\documentclass[lineno]{jfm}

\usepackage{graphicx}
\usepackage{newtxtext}
\usepackage{newtxmath}
\usepackage{natbib}
\usepackage{hyperref}
\usepackage{tikz} 
\usepackage{subcaption}

\newcommand{\RomanNumeralCaps}[1]

\title{Achieving Optimal Locomotion using Self-Generated Waves}

\author{D O'Donovan\aff{1}
  \corresp{\email{daire.odonovan@ucdconnect.ie}},
  MD Bustamante\aff{1},
  O Devauchelle\aff{2}
 \and GP Benham\aff{1}}

\affiliation{\aff{1}School of Mathematics and Statistics, University College Dublin, Belfield, Dublin 4, Ireland
\aff{2}Université de Paris, Institut de Physique du Globe de Paris, CNRS, F-75005 Paris, France}

\begin{document}
\maketitle

\begin{abstract}
    An oscillating body floating at the water surface produces a wave-field of self-generated waves. When the oscillation induces a difference in fore-aft wave amplitude squared, these self-generated waves can be used as a mechanism to propel the body horizontally across the surface \citep{LONGUETHIGGINS1964}. The optimisation of this wave-driven propulsion (WDP) is the interest of this work. To study the conditions necessary to produce optimal thrust we will utilise a shallow water set-up where a periodically oscillating pressure source acts as the body. In this framework, an expression for the thrust is derived by relation to the aforementioned difference in fore-aft amplitude squared. The conditions on the source for maximal thrust are explored both analytically and numerically in two optimal control problems. The first case is where a bound is imposed on the norm of the control function to regularise it. Secondly, a more physically motivated case is studied where the power injected by the source is bounded. The body is permitted to have a drift velocity $U$. When scaled with the wave speed $c$, the dimensionless velocity $v=U/c$ divides the study into subcritical, critical and supercritical regimes and the optimal conditions are presented for each. The result in the bounded power case is then used to demonstrate how the modulation of power injected can slowly change the cruising velocity from rest to supercritical velocities.
    
\end{abstract}

\section{Introduction}

Horizontal propulsion along the water surface is seen in the natural world with examples including various insects (e.g water striders and whirligig beetles) \citep{Bush2006review} as well as ducks \citep{Yuan2021ducks}. Motivated by a curiosity to understand the underpinning fluid mechanics describing the actions of these creatures, various studies have been conducted such as high speed imaging of insects like the water strider \citep{Hu2003water_strider,Steinmann2021} and the stricken honeybee \citep{Roh2019bee}. Inspired by these creatures, robots that mimic their movements have been developed, with \textit{Robostrider} \citep{Hu2003water_strider} in the case of the water strider and \textit{SurferBot} in the case of the honeybee \citep{Rhee2022surferbot}. In a similar vein to such movements we can achieve horizontal locomotion by riding waves produced by gunwale bobbing on a canoe \citep{Benham2022gunwale}.

The water strider was first thought to use Wave-Driven Propulsion (WDP), but the observation of subsurface vortex generation \citep{Hu2003water_strider} shifted our understanding to the use of WDP in combination with vortex-driven propulsion (VDP) \citep{BUHLER_2007, Steinmann2018,GAO_FENG_2011,Steinmann2021}. We note that although multiple inertial propulsion mechanisms can be used including vortices (VDP), waves (WDP) or a combination of both (as in the water strider case), in this study we will focus on optimising thrust using WDP only.  

In the case of the honeybee trapped on the water surface, it beats its wings to create an asymmetry in the fore-aft surface wave-field which is indicative of the use of radiation stresses described by \citet{LONGUETHIGGINS1964} where the former later applied these ideas to propel a small craft \citep{Longuet-Higgins1977meanforces}.  Radiation stress is defined as proportional to the difference in the square of the fore-aft amplitudes generating momentum and hence thrust. Mathematically in the shallow limit the thrust force is \citep{LONGUETHIGGINS1964}
\begin{equation}\label{eq:L-H_reference}
    F_T = \frac{3}{4}\rho g\Big[h^2\Big]^{+}_{-},
\end{equation}
where $h$ is the wave amplitude, $\rho$ is density, $g$ is acceleration due to gravity and $+$,$-$ denotes the fore and aft waves respectively. 
Radiation stress can be used to find the thrust generated by waves to propel bodies on a vibrating surface \citep{Ho2023capillary_surfers,Pucci2015floating_drops,Barotta2023spinners} as well as a body producing its own waves in WDP \citep{Longuet-Higgins1977meanforces,Benham2024propulsion}.
Motivated by the latter, we will characterise the thrust due to WDP by starting from the wave equation and  recovering the difference in fore-aft amplitude squared. To reduce complexity, we will work in the shallow water approximation and focus only on surface waves ignoring any subsurface effects.
We also need to characterise the body which we will be studying. Rather than a rigid body on the surface  \citep{Benham2024propulsion}, we will keep things general and consider a pressure region influencing the surface similar to the characterisation of an air-cushion vehicle \citep{Doctors_1972}.

Once we have built the mathematical problem, we can look at posing an optimal control problem to maximise the thrust. Optimisation studies are motivated not only by understanding the natural world with examples including ducklings wave-riding in their mother's wake \citep{Yuan2021ducks} but also for industrial applications. Optimising WDP can lead to improved fuel efficiency of boats by extracting propulsive energy from the waves \citep{Bockmann2018foils}. Meanwhile in sports, optimisation of WDP can be used to gain an advantage by using opponents’ waves to improve propulsion \citep{Tuck1998OptimumHS,Dode2022}. 
This paper will consider an optimal control problem from a different perspective compared to the examples in \citet{Yuan2021ducks} and \citet{Doctors_1997}. Rather than minimising wave resistance, we are interested in maximising the thrust due to WDP. The only constraints on the problem are the shallow water setup and appropriate bounds on the pressure source.

The optimal control problem is divided into multiple cases according to the value of the drift velocity $U$. The case where the drift velocity $U$ is zero will be studied in Sections \ref{sec:second_section} and \ref{sec:startup}. 
This will simplify the expressions to allow us to focus on the concepts such as deriving the thrust and the power. 
Following the completion of the start-up case, Section \ref{sec:subcritical} will be concerned with the case where the drift velocity is less than the wave speed (subcritical). 
This section will introduce the application of a Galilean transformation to operate in the rest frame of the body. The case where the body moves at the same velocity as the waves emitted (critical) will follow in Section \ref{sec:critical}. 
The intuition will be useful when considering the case where the body is travelling faster than the wave speed (supercritical) in Section \ref{sec:supercritical} requiring us to think carefully about the left and right wave since they are both left-travelling after the transformation. 
This will complete the template of velocity regimes from which we can study the injection of power necessary to accelerate from rest to supercritical velocities in Section \ref{sec:quasiperiodic}, given that the acceleration is small enough to maintain the periodic assumptions that are made. 
Finally, Section \ref{sec:Discussion} discusses the scope and possible improvements of this work. 

\section{1D shallow water waves due to a source}\label{sec:second_section}

Given a single body of length $L$ propelling itself forward using WDP, we wish to study the conditions to produce maximal forward thrust. We will consider a surface region where there is a pressure disturbance induced by the motion of the body. To begin, we will consider zero horizontal velocity $U$ of the body. This will be referred to as the start-up phase where a body is beginning to propel itself forward from rest. Following this, the horizontal velocity will be introduced in three different regimes: subcritical velocities ($0<v<1$), critical velocities ($v=1$) and finally supercritical velocities ($v>1$), where $v=U/c$ is the drift velocity divided by the wave speed $c$. 

\begin{figure}
\centering  
\begin{tikzpicture}[scale=1.1]  
    \node at (-10,0) {\includegraphics[width=0.855\textwidth]{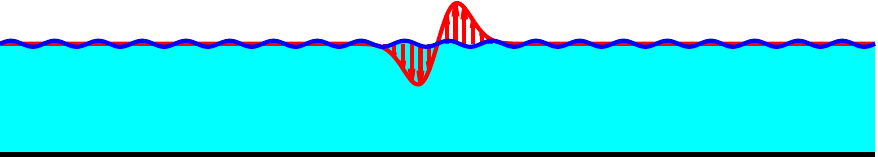}};  
    \node at (-15.5,1.1) {$z$};  
    \node at (-8.6,0.7) {\color{red}{$p=P(x,t)$}};  
    \draw[line width=2,<->,red] (-10.5,-0.5) -- (-9.5,-0.5);  
    \node at (-10,-0.25) {\color{red}$L$};  
    \draw[line width=1,->,black] (-15.25,-0.9474) -- (-15.25,1);  
    \draw[line width=1,->,black] (-15.25,-0.93) -- (-4.73,-0.91); 
    \node at (-4.6, -1.1) {$x$};
    \draw[line width=2,<->,blue] (-15.5,-0.9) -- (-15.5,0.4); 
    \node[blue] at (-15.8,-0.25) {$H$};  
    \draw[line width=2,->,blue] (-12,0.8) -- (-12,0.5);  
    \draw[line width=2,->,blue] (-12,0.05) -- (-12,0.35); 
    \node[blue] at (-12.6,0.7) {$h(x,t)$};  
    \draw[line width=1,->,black] (-8.25,1.2) -- (-7.25,1.2); 
    \node at (-7,1.2) {$U$};  
\end{tikzpicture}  
\caption{Diagram of the set-up where there is a pressure source $P(x,t)$, over a region of length $L$, acting on the surface of the shallow water, with depth $H$. This induces waves on the fluid surface, denoted by $h(x,t)$, resulting in locomotion with a drift velocity $U$.}  \label{fig:Setup}
\end{figure}
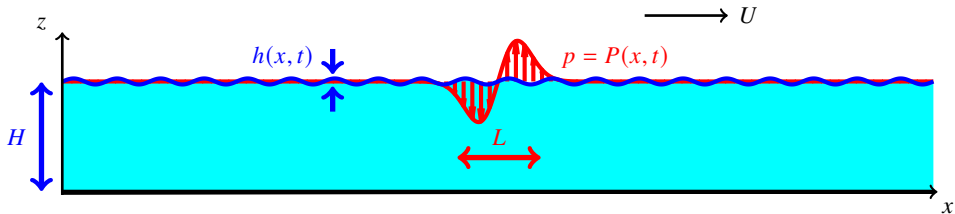

To add a simplification which allows us to focus on the surface of the water, we will operate within the shallow water approximation and a one-dimensional set-up as shown in figure \ref{fig:Setup}. A pressure source $P(x,t)$ is applied at the surface of an otherwise undisturbed body of shallow water where the fluid is in the region $0\leq z\leq H$ in the steady state.
The source induces a perturbation so the surface of the water is at $z=H+h(x,t)$, where $h(x,t)\ll H$. The shape of the surface perturbed by the pressure is encapsulated in $h(x,t)$. Unlike the honeybee and the water strider which produce gravity-capillary waves, here we focus on gravity waves and ignore the effect of surface tension. 

As described in Appendix \ref{appendix_derive_shallow_wave_eq}, the perturbed wave field $z=h(x,t)$ satisfies the one-dimensional wave equation,
\begin{equation}\label{eq:dimensional_wave}
    \frac{\partial^2h}{\partial x^2} - \frac{1}{c^2}\frac{\partial^2h}{\partial t^2} = -Q(x,t), \hspace{1cm} -\infty < x < \infty,
\end{equation}
where $c=\sqrt{gH}$ is the wave speed and $Q(x,t) = P_{xx}(x,t)/\rho g$ is the pressure source term. The source is defined to be non-zero in the region $x\in\left[-\frac{L}{2},\frac{L}{2}\right]$ and zero outside this. Since this investigation is concerned with a single body undergoing periodic oscillations, with a frequency $\omega$, the variables will be rescaled as follows:
\begin{align}\label{eq:freq_scaling}
    h, x \sim \frac{c}{\omega},& & t\sim \frac{1}{\omega}, & &  Q \sim \frac{ \omega}{c}.
\end{align}
After considering these scalings, we arrive at the dimensionless wave equation with a source,
\begin{equation}\label{eq:dimensionless_wave}
        \frac{\partial^2h}{\partial x^2} - \frac{\partial^2h}{\partial t^2} = -Q(x,t), \hspace{1cm} -\infty < x < \infty.
\end{equation}
In dimensionless form, the source $Q$ is defined to be non-zero on the interval $\left[-\frac{l}{2},\frac{l}{2}\right]$, where $l$ is the dimensionless body length $l = L\omega/c$.  A periodically oscillating source and wave field are assumed, since there is a single body and the surface is otherwise undisturbed. Therefore, the source and wave field can be decomposed as follows:
\begin{align}\label{eq:v0_periodic_decomp}
    h(x,t) = \mathrm{Re}\left[\hat{h}(x)e^{it}\right], & & Q(x,t) = \mathrm{Re}\left[\hat{Q}(x)e^{it}\right].
\end{align}
The above expressions in \eqref{eq:v0_periodic_decomp} are used to reduce the PDE for $h(x,t)$ in \eqref{eq:dimensionless_wave} to an ODE for the complex function $\hat{h}(x)$,
\begin{equation}\label{eq:v0_ode}
    \hat{h}''(x)+\hat{h}(x) = -\hat{Q}(x).
\end{equation}
The ODE has a general solution that is calculated using Green's functions,
\begin{equation}\label{eq:v0_ode_gen_sol}
    \hat{h}(x) = Ae^{ix} + Be^{-ix} + \frac{i}{2}\int^{x}_{-\frac{l}{2}}\hat{Q}(X)\left(e^{i\left(x-X\right)} - e^{i\left(X-x\right)}\right)\,\mathrm{d}X,
\end{equation}
where $A,B\in\mathbb{C}$ are constants to be found. We note that although there is a dependence on $x$ in the integral, for $x>\frac{l}{2}$ the upper bound will be $\frac{l}{2}$ since $\hat{Q}(X>\frac{l}{2})=0$. Likewise for $x<-\frac{l}{2}$ the integral is zero since $Q(X<-\frac{l}{2}) = 0$. The following Sommerfeld boundary conditions will be used, where $k_{\pm}$ is the wavenumber:  
\begin{align}\label{eq:v0_radiative_bc}
    \hat{h}'\left(\pm\frac{l}{2}\right) = \mp ik_{\pm}\hat{h}\left(\pm\frac{l}{2} \right),
\end{align}
and $k_{\pm}=1$ for the start-up case. This is equivalent to imposing that waves only radiate away from the source. Through application of the boundary conditions, $A$ and $B$ are found to be
\begin{align}\label{eq:v0_vals_for_constants}
    A = -\frac{i}{2}\int^{\frac{l}{2}}_{-\frac{l}{2}}\hat{Q}(X)e^{-iX}\,\mathrm{d}X, && B = 0,
\end{align}
and the resulting solution for the wave-field is
\begin{equation}\label{eq:v0_h_sol}
    \hat{h}(x) = -\frac{i}{2}\int^{\frac{l}{2}}_{x}\hat{Q}(X)e^{i\left(x-X\right)}\,\mathrm{d}X - \frac{i}{2}\int^{x}_{-\frac{l}{2}}\hat{Q}(X)e^{i\left(X-x\right)}\,\mathrm{d}X.
\end{equation}
This solution is therefore composed only of a left-travelling wave for $x<-\frac{l}{2}$ and a right travelling wave for $x>\frac{l}{2}$.

The problem we aim to solve is to find a $\hat{Q}$ such that it gives maximal thrust. Therefore, an expression for the thrust is required. The thrust force is not immediately obvious to define but can be derived through manipulation of the wave equation \eqref{eq:dimensionless_wave} to recover the ``radiation stress'' as described by \citet{LONGUETHIGGINS1964}, i.e. an asymmetrical wave-field resulting in a difference in fore-aft amplitude that generates thrust. Multiplying \eqref{eq:dimensionless_wave} by $h_x$ and taking an integral with respect to $x$ (see Appendix \ref{ap:thrust_derive}), the radiation stress result can be recovered where the difference in fore-aft amplitude squared ($|\hat{h}|^2$) produces a thrust. As such we derive the following equivalence:
\begin{equation}\label{eq:v0_L-H_result}
    -\frac{1}{2}\left[k|\hat{h}|^2\right]^{+}_{-} = \langle\hat{Q},\hat{h}'\rangle,
\end{equation}
where the $+$ and $-$ symbols in \eqref{eq:v0_L-H_result} represent values for $x$ that are on the right and left hand side of the $[-\frac{l}{2},\frac{l}{2}]$ region respectively and $\langle\cdot,\cdot\rangle$ is the time-averaged inner product defined as 
\begin{equation}\label{eq:inner_product_defn}
    \langle f,g\rangle \coloneq \frac{1}{T}\int^{T}_{0}\int^{\frac{l}{2}}_{-\frac{l}{2}} \textrm{Re}\left[f(x)e^{it}\right]\,\textrm{Re}\left[g(x)e^{it}\right]\,\textrm{d}x\,\textrm{d}t = \frac{1}{4}\int^{\frac{l}{2}}_{-\frac{l}{2}} f\,g^{*} + f^{*}g\,\textrm{d}x.
\end{equation}
The left hand side of \eqref{eq:v0_L-H_result} has included differing fore-aft wavenumbers $k_{\pm}$, but in the start-up case $k_+=k_- = 1$. This will be more important when velocity is introduced to apply a correction due to the Doppler shift introduced through operating in the moving frame. Relating back to the radiation stress, we interpret \eqref{eq:v0_L-H_result} as an equivalence relation between the force injected and the resulting asymmetrical wave-field that induces thrust. It is noted that there is a different prefactor compared to \eqref{eq:L-H_reference} which can easily be fixed through scaling the equations by $3/2$, but this will not be done since it has no effect on the work to follow. Hence, motivated by \eqref{eq:v0_L-H_result} the time-averaged thrust is taken to be 
\begin{equation}\label{eq:v0_thrust_with_quarter}
    \bar{F}_T\left(\hat{Q}(x)\right) \coloneq \int^{\frac{l}{2}}_{-\frac{l}{2}}\overline{Qh_x}\,\mathrm{d}x = \langle\hat{Q},\hat{h}'\rangle.
\end{equation}

Some constraints will be required in the optimal control problem. One case that will be investigated is when the time-averaged power is bounded. To derive an expression for the power, we begin with the energy of the wave as the sum of the dimensionless kinetic and potential energy integrated over the whole domain
\begin{equation}\label{eq:v0_energy}
    E(t) = \int^{+}_{-} \left(\frac{1}{2}h_{t}^2+\frac{1}{2}h_x^2 \right)\,\mathrm{d}x.
\end{equation}
Power is given as $\mathrm{Pow} = \frac{\mathrm{d}E}{\mathrm{d}t}$. Since we are working with periodic oscillations, the time-averaged power is $\overline{\mathrm{Pow}} =E(T) - E(0) = 0$, which means the energy is conserved over a period. The right hand side of \eqref{eq:v0_energy} can be differentiated with respect to time and \eqref{eq:dimensionless_wave} is substituted and then time-averaged to generate another equivalence relationship (see Appendix \ref{ap:power_derive}):
\begin{equation}\label{eq:power_equality}
    \int^{\frac{l}{2}}_{-\frac{l}{2}} \overline{Qh_t}\, \mathrm{d}x =-\left[\overline{h_th_x}\right]^{+}_{-}. 
\end{equation}
The average power injected by the pressure source on the left-hand side of \eqref{eq:power_equality} is related by equality to the average power radiated at the boundaries of the pressure source on the right hand side of \eqref{eq:power_equality}. Since this is an idealised system where no power inputted is lost due to external effects, this is expected. Motivated by \eqref{eq:power_equality} we take the time-averaged power injected to be
\begin{equation}\label{eq:v0_power}
    \overline{\mathrm{Pow}} \coloneq \int^{\frac{l}{2}}_{-\frac{l}{2}} \overline{Qh_t}\, \,\mathrm{d}x= \langle\hat{Q},i\hat{h}\rangle.
\end{equation}
Furthermore, we can substitute \eqref{eq:v0_periodic_decomp} into the right hand side of \eqref{eq:power_equality} to show
\begin{equation}
    \frac{1}{2}\left[k_{+}|\hat{h}_{+}|^2 + k_{-}|\hat{h}_{-}|^2\right] = \langle\hat{Q},i\hat{h}\rangle. 
\end{equation}
As such, while the thrust is the difference in fore-aft amplitude squared, the power is the sum.

Now, we can proceed with posing the optimal control problem. In the work so far, the body is taken to be at rest. The thrust would introduce movement itself but, for now, we will restrict the problem to that of ``start-up,'' with the horizontal drift velocity close to zero, $U\approx0$.

\section{Optimal locomotion on start-up}\label{sec:startup}

We wish to pose an optimal control problem to maximise the thrust as it is defined in \eqref{eq:v0_thrust_with_quarter}. The control function $\hat{Q}(x)$ is in the expression for $\hat{h}(x)$ in \eqref{eq:v0_h_sol} which implicitly enforces that the wave ODE \eqref{eq:v0_ode} and its boundary conditions \eqref{eq:v0_radiative_bc} must be satisfied. Some sort of regularisation to the problem is required. In this paper we will use two different constraints. Firstly we will bound the time-averaged norm of the control,
\begin{equation}\label{eq:v0_norm_expression}
    \langle\hat{Q},\hat{Q}\rangle \leq \varepsilon,
\end{equation}
for some dimensionless $\varepsilon>0$\footnote{We define $\varepsilon = \hat{\varepsilon}c/\omega$, $\delta = \hat{\delta}\omega/\rho gH^{2}c$ where $\hat{\varepsilon},\hat{\delta}$ are dimensional bounds.}. Secondly, we will bound the time-averaged power, as it is seen in \eqref{eq:v0_power},
\begin{equation}\label{eq:power_bound}
    \langle\hat{Q},i\hat{h}\rangle \leq \delta,
\end{equation}
for some dimensionless $\delta>0$. It is worth noting that, neither constraint requires that $\hat{Q}$ be continuous at the boundary and jumps may be present at $x=\pm\frac{l}{2}$. 

Along with an analytical approach, a numerical optimisation approach is taken to validate the results using the Ipopt solver that is part of the JuMP package in Julia \citep{Lubin2023jump}. To pose the problem numerically, the inputs are divided into variables, objective and constraints. There are four arrays of variables to represent the real and imaginary parts of $\hat{Q}$ and $\hat{h}$ which are discretised in space. The thrust \eqref{eq:v0_thrust_with_quarter}, discretised using the trapezium rule, is the objective function we wish to maximise. As for the constraints, these are the boundary conditions \eqref{eq:v0_radiative_bc}, the discretised ODE for $\hat{h}(x)$ \eqref{eq:v0_ode}, and either the bounded norm of $\hat{Q}$, or the bounded power. In the interest of comparing the analytical and numerical work, we take $\varepsilon,\delta=1$. These will be dealt separately in the subsections that follow.

\subsection{Case 1 - Bounded Norm}\label{subsec:v0_nrm}

To be physically plausible, the source $Q$ can be regularised by bounding the $L_2$ norm which will provide us with well behaved solutions for $\hat{Q}$. In \eqref{eq:v0_norm_expression} the value $\varepsilon=1$ is taken to compare the analytical and numerical approaches. The variational calculus approach in Appendix \ref{norm_appendix} results in the following condition to be an optimal solution,
\begin{equation}\label{eq:v0_norm_ode_sol}
    \mathcal{L}_0\hat{Q} = 0,
\end{equation}
where the (1D Helmholtz) differential operator $\mathcal{L}_0 = \partial_{xx} + \mathbb{I}$, applied to $\hat{Q}$ implies wave-like solutions,
\begin{equation}\label{eq:v0_norm_sol}
    \hat{Q}(x) = D_{\mathrm{L}}\,e^{ix} +D_{\mathrm{R}}\,e^{-ix},   
\end{equation}
where $D_{\mathrm{L}},D_{\mathrm{R}}\in\mathbb{C}$. In Appendix \ref{norm_appendix}, we go a step further and find the constants $D_{\mathrm{L}}$ and $D_{\mathrm{R}}$ in terms of $\lambda$, the Lagrange multiplier, which is also found. The constraint on the modulus squared of the constants is
\begin{equation}\label{eq:v0_nrm_cnsts}
    \begin{aligned}
        |D_{\mathrm{L}}|^2 = 4\left(l+\frac{l\sin^2(l)}{(2\lambda-l)^2}+\frac{2\sin^2(l)}{2\lambda-l}\right)^{-1}, && && |D_{\mathrm{R}}|^2 = |D_{\mathrm{L}}|^2\frac{\sin^2(l)}{(2\lambda-l)^2},
    \end{aligned}
\end{equation}
where $\lambda$ is
\begin{equation}\label{eq:v0_lambda}
    \lambda_{\pm} = \pm\frac{\sqrt{l^2- \sin^2(l)}}{2}.
\end{equation}
The $(\pm)$ in the expression for $\lambda$ corresponds to the choice of thrust direction. To move in the positive direction (right) corresponds to taking the negative part in $\lambda$, thus maximising positive thrust. Since $\lambda$ is the Lagrange multiplier, $\lambda\in\mathbb{R}$ is required. We have the condition
\begin{equation}\label{eq:v0_real_condition_nrm}
    l^2 - \sin^2(l)>0,
\end{equation}
which is true for all real values of $l$, so $\lambda$ is always real. Hence, in \eqref{ap_eq:A-Brelation} we note $D_{\mathrm{L}}$ and $D_{\mathrm{R}}$ are related via multiplication by a real number meaning the two waves are in phase.

\begin{figure}
    \centering
    \begin{subfigure}[b]{0.45\textwidth}
        \centering
        \includegraphics[width=\textwidth]{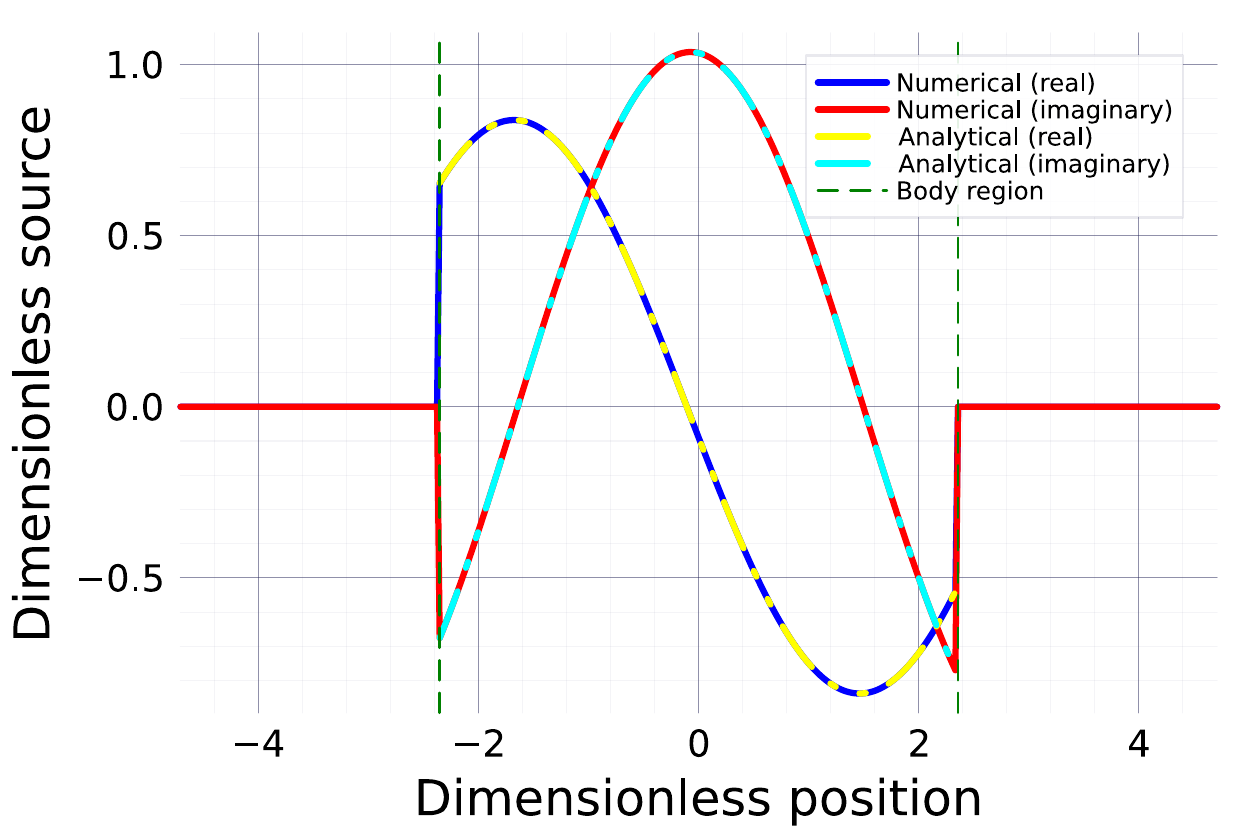}
        \caption{}
        \label{v0_norm_Q_plot}
    \end{subfigure}
    \begin{subfigure}[b]{0.45\textwidth}
        \centering
        \includegraphics[width=\textwidth]{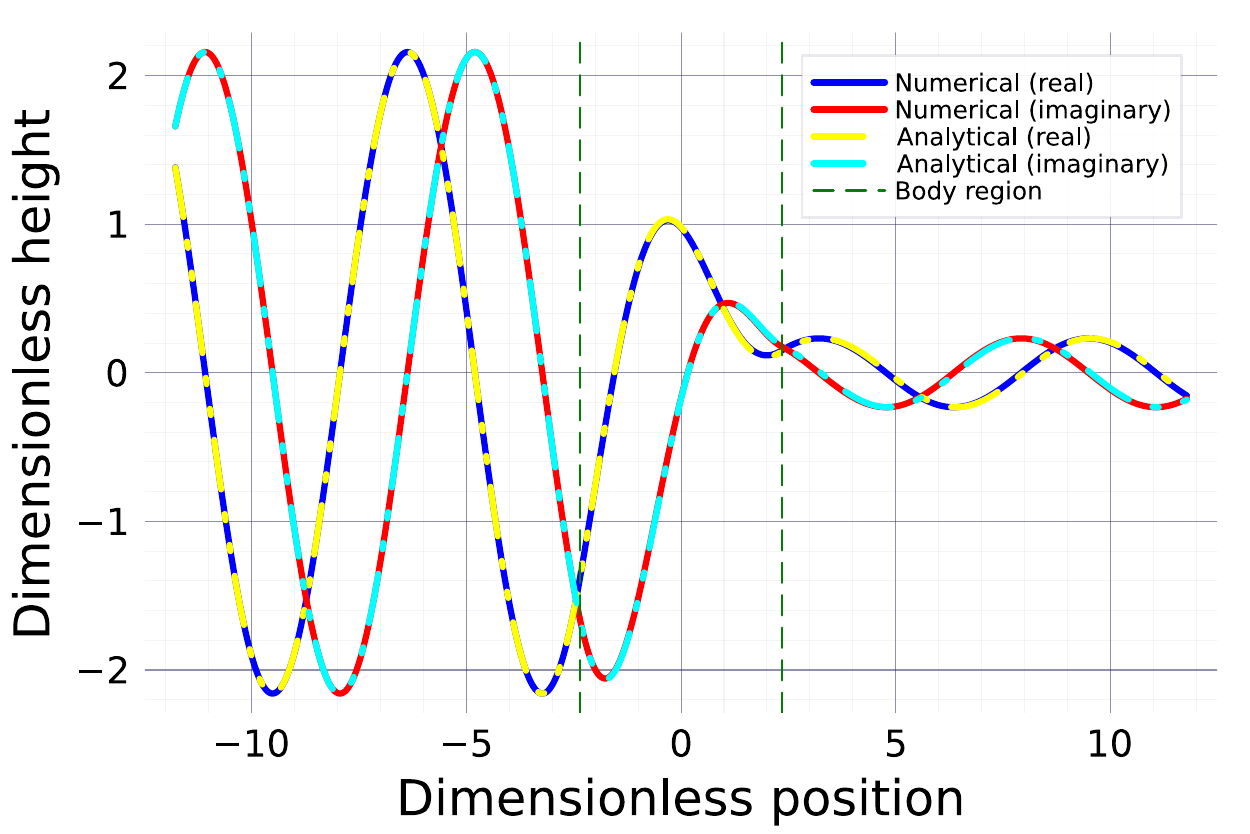}
        \caption{}
        \label{v0_norm_h_plot}
    \end{subfigure}
    \caption{(a) An example of a source, $\hat{Q}(x)$, that results in the optimal thrust under the bounded norm constraint where $v=0$ and $l=3\pi/2$. (b) The corresponding wave-field, $\hat{h}(x)$ under the same conditions.}
    \label{fig:v0_norm}
\end{figure}

A periodic source $\hat{Q}$ that is composed of a left travelling and/or right travelling waves of this form will produce optimum thrust. There are infinitely many such solutions which differ by an arbitrary choice of phase $\mathrm{Arg}(D)$. An example of one such $\hat{Q}(x)$, found numerically, is displayed in figure \ref{fig:v0_norm} along with the resulting wave-field $\hat{h}(x)$. In figure \ref{v0_norm_h_plot} we see a larger wave on the left hand side which corresponds to positive thrust and hence start-up locomotion in the right-travelling direction is achieved.

The expressions for $|D_{\mathrm{L}}|^2$ and $|D_{\mathrm{R}}|^2$ in \eqref{eq:v0_nrm_cnsts} are validated numerically for a given $l$. Using \eqref{ap_eq:DL_DR_to_numerics} allows us to use the right and left waves to relate $|D_{\mathrm{L}}|^2,|D_{\mathrm{R}}|^2$ to the information we have on $\hat{Q}$ from the numerical result which is one of infinite solutions that satisfy
\begin{equation}\label{eq:finding_DL_DR}
    \begin{aligned}
        \left|\frac{1}{2}\int^{\frac{l}{2}}_{-\frac{l}{2}}\hat{Q}(X)e^{-iX}\,\mathrm{d}X\right|^2 = \lambda^2|D_{\mathrm{L}}|^2,
        && &&
        \left|\frac{1}{2}\int^{\frac{l}{2}}_{-\frac{l}{2}}\hat{Q}(X)e^{iX}\,\mathrm{d}X \right|^2 = \lambda^2|D_{\mathrm{R}}|^2.
    \end{aligned}
\end{equation}
We note that $\lambda^2|D_{\mathrm{L}}|^2$ and $\lambda^2|D_{\mathrm{R}}|^2$ are related to the amplitude squared of the left and right travelling waves ($|\hat{h}_{\mathrm{L}}|^2,|\hat{h}_{\mathrm{R}}|^2$) respectively.
For $l=3\pi/2$, the relative error between the analytical and numerical expressions was found to be $0.003\%$ for $\lambda|D_{\mathrm{L}}|^2$ and $0.014\%$ for $\lambda^2|D_{\mathrm{R}}|^2$. 
Given the aforementioned arbitrary choice of phase, the numerical scheme will converge to one of an infinite set of solutions. If we use \eqref{ap_eq:DL_DR_to_numerics}, we find the numerical values for $D_\mathrm{L}$ and $D_{\mathrm{R}}$ and use them in conjunction with \eqref{eq:v0_norm_sol} to demonstrate correspondence between the numerical and analytical results in figure \ref{fig:v0_norm}.  

\begin{figure}
    \centering
    \includegraphics[width=0.5\linewidth]{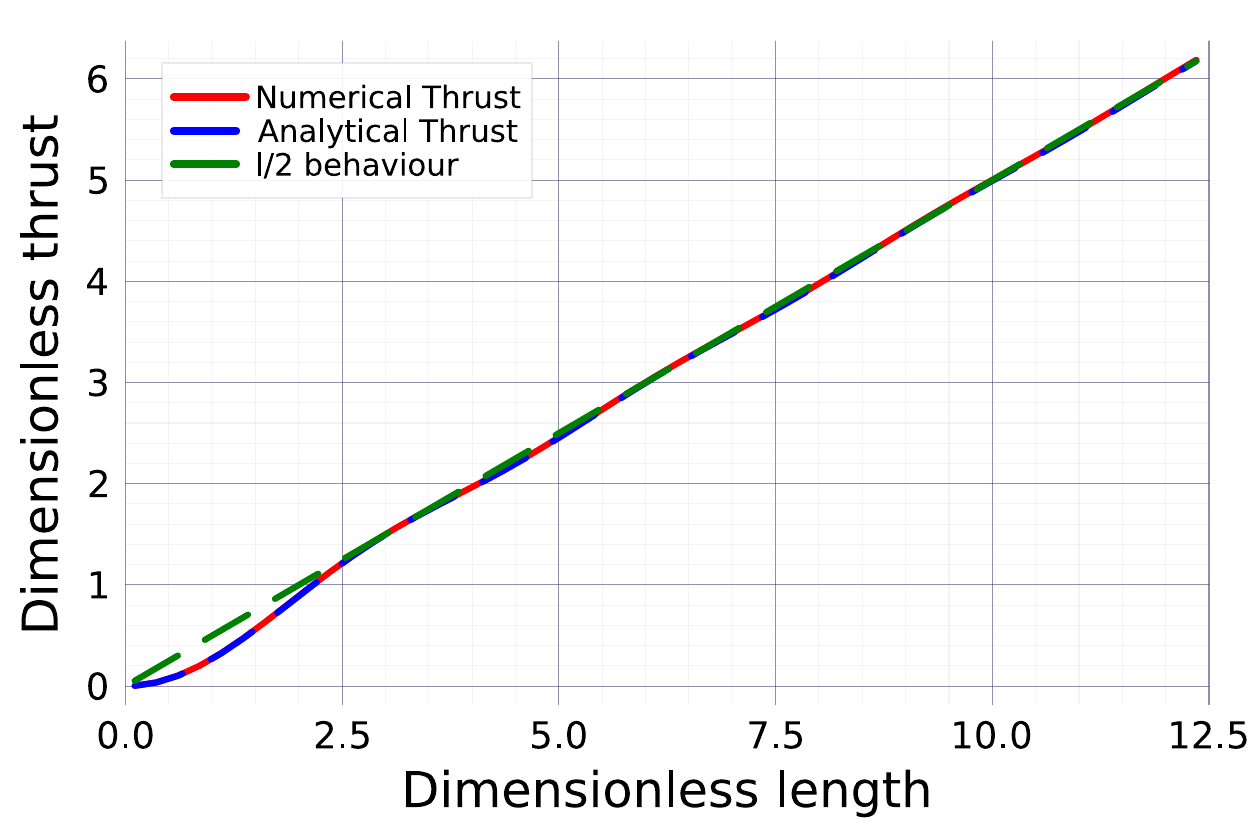}
    \caption{Plot of the time-averaged thrust $\bar{F}_T$ vs the dimensionless length scale $l$ resulting from the bounded norm optimisation on start-up $v\approx0$. It can also be seen that as $l$ increases, the dimensional thrust tends to scale with $\frac{l}{2}$ which is the same behaviour found in \eqref{eq:norm_force_v0} for $l\gg1$ }
    \label{fig:v0_lsweep}
\end{figure}

The dependence of the expressions on the length scale $l$, which is related to $\omega$, motivates that we investigate how the resulting thrust depends on $l$. The thrust can be written (see Appendix \ref{norm_appendix}) in terms of the constants that have been found, 
\begin{equation}\label{eq:norm_force_v0}
    \bar{F}_T = \frac{1}{4}\left(2\lambda^2|D_{\mathrm{L}}|^2 - 2\lambda^2|D_{\mathrm{R}}|^2\right).
\end{equation}
Plotting the force that results from the numerical optimisation, we see that it increases with $l$ (figure \ref{fig:v0_lsweep}), which is a property we will return to in Section \ref{sec:subcritical}.  

\subsection{Case 2 - Bounded power}\label{subsec:v0_power}

Rather than bounding the norm of $Q$, let us instead bound the power \eqref{eq:v0_power}. A body on the water propelling itself through vertical oscillations has control over the power it inputs to generate thrust. Mathematically, the power is bounded by $\overline{\textrm{Pow}} \leq \delta$, where $\delta=1$ is chosen for the following work.  The variational calculus approach in Appendix \ref{power_appendix} is taken and the optimal condition is a first order homogeneous ODE,
\begin{equation}\label{eq:v0_eigenval_prob}
    \left(\partial_{x} + i\lambda\right)\left(-\frac{i}{2}\int^{\frac{l}{2}}_{-\frac{l}{2}}\hat{Q}(X)e^{i(X-x)}\,\mathrm{d}X - \frac{i}{2}\int^{\frac{l}{2}}_{-\frac{l}{2}}\hat{Q}(X)e^{i(x-X)}\,\mathrm{d}X\right) = 0,
\end{equation}
which results in two solutions depending on the eigenvalue, $\lambda\in\mathbb{R}$, which again corresponds to the Lagrange multiplier in the variational calculus. The two solutions for $\lambda$ will be denoted as $\lambda_1$ and $\lambda_2$ and the corresponding solutions to the ODE \eqref{eq:v0_eigenval_prob} are given as follows:
\begin{equation}\label{eq:v0_option1}
	C_{\mathrm{L}}\,e^{ix} + C_{\mathrm{R}}\,e^{-ix} = C\,e^{ix}\,\,\,\,\,\,\,\,\textrm{for}\,\, \lambda_{1}=-1,
\end{equation}
\begin{equation}\label{eq:v0_option2}
	C_{\mathrm{L}}\,e^{ix} + C_{\mathrm{R}}\,e^{-ix} = C\,e^{-ix}\,\,\,\,\,\,\,\,\textrm{for}\,\, \lambda_{2} = 1,
\end{equation}
where $C_{\mathrm{L}},C_{\mathrm{R}} \in\mathbb{C}$. This gives us the option of a purely left or purely right travelling wave and to illuminate this fact, the left and right wave amplitudes have been labelled $C_{\mathrm{L}}$ and $C_{\mathrm{R}}$ respectively. The constants are found by equating coefficients to be 
\begin{align}\label{eq:v0_equated_coeffs}
    C_{\mathrm{L}} = -\frac{i}{2}\int^{\frac{l}{2}}_{-\frac{l}{2}}\hat{Q}(X)e^{-iX}\,\mathrm{d}X, & & C_{\mathrm{R}} = -\frac{i}{2}\int^{\frac{l}{2}}_{-\frac{l}{2}}\hat{Q}(X)e^{iX}\,\mathrm{d}X.
\end{align}
Similar to the bounded norm case, we can note that $C_{\mathrm{L}}$ and $C_{\mathrm{R}}$ correspond the fore-aft wave heights respectively. Before making a choice of $\lambda$ and discussing that solution, we recall that the bounded power \eqref{eq:v0_power}, when written in terms of the above constants, gives a condition on $|C_{\mathrm{L}}|^2$ and $|C_{\mathrm{R}}|^2$, 
\begin{equation}\label{eq:v0_constants_relation}
    \frac{1}{4}\left(2\left|C_{\mathrm{L}} \right|^2 + 2\left|C_{\mathrm{R}}\right|^2\right) = 1.
\end{equation}
To achieve a positive thrust, a purely left travelling wave is chosen ($\lambda=-1$), meaning $C_{\mathrm{L}} = C$ and $C_{\mathrm{R}} = 0$ (see Appendix \ref{power_appendix}). Using \eqref{eq:v0_constants_relation} the condition for optimal thrust is  
\begin{align}\label{eq:v0_amplitude_cond}
    \left|-\frac{i}{2}\int^{\frac{l}{2}}_{-\frac{l}{2}}\hat{Q}(X)e^{-iX}\,\mathrm{d}X\right|^2 = 2, && \mathrm{and} && \left|-\frac{i}{2}\int^{\frac{l}{2}}_{-\frac{l}{2}}\hat{Q}(X)e^{iX}\,\mathrm{d}X\right|^2 = 0.
\end{align}
These equations are two linear constraints on the function $\hat{Q}$, constituting thus an underdetermined system. Therefore there are infinitely many possible solutions that satisfy this. 
To verify that there is indeed a positive thrust, the conditions \eqref{eq:v0_amplitude_cond} are substituted into the thrust formula \eqref{eq:v0_thrust_with_quarter} to give
\begin{equation}\label{eq:v0_thrust_expectation}
    \bar{F}_T = 1.
\end{equation}
This makes physical sense, since to get a motion in a right travelling direction, a left travelling wave would be required. The absence of a wave in the rightward direction agrees with this intuition since a wave on the right hand side would reduce the difference in fore-aft amplitude which would reduce the radiation stress and hence reduce the thrust by wasting some power injected. 

\begin{figure}
    \centering
    \begin{subfigure}[b]{0.45\textwidth}
        \centering
        \includegraphics[width=\textwidth]{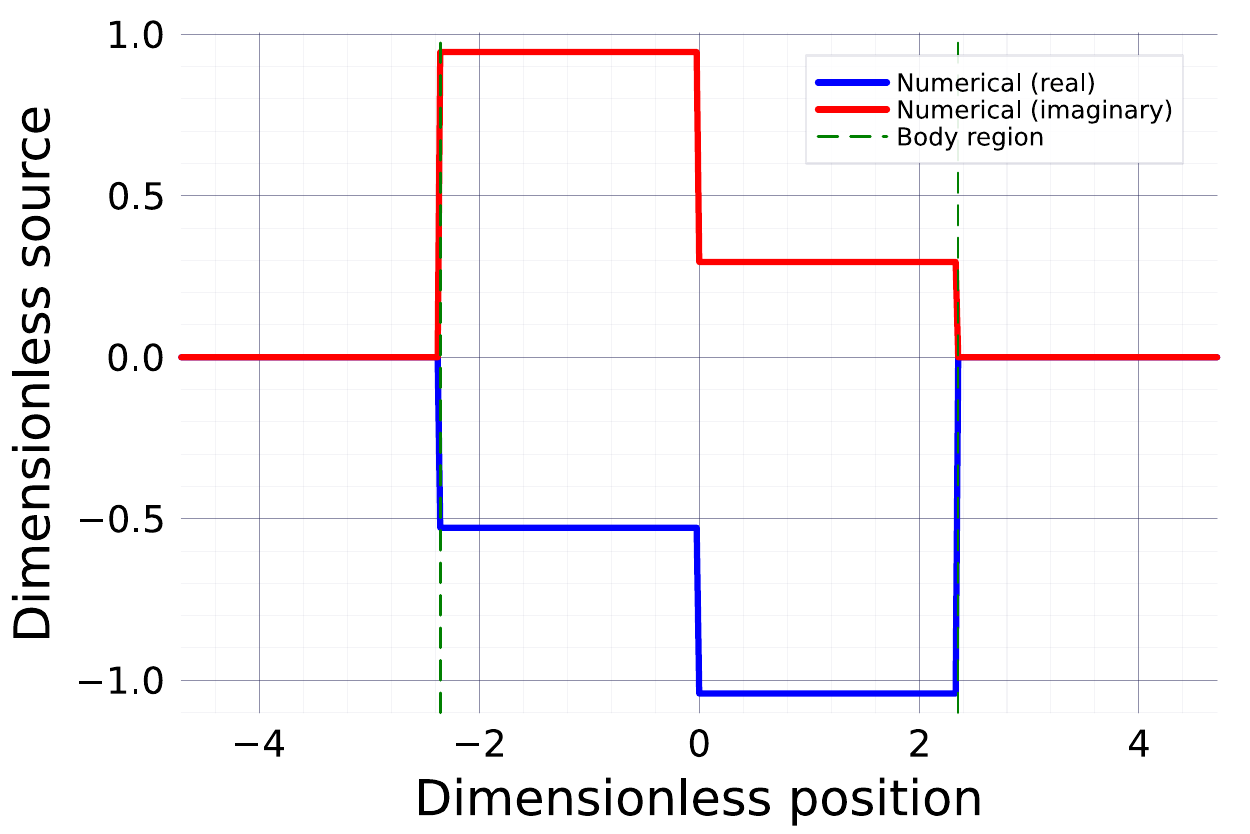}
        \caption{}
        \label{v0_pwr_q_plot}
    \end{subfigure}
    \begin{subfigure}[b]{0.45\textwidth}
        \centering
        \includegraphics[width=\textwidth]{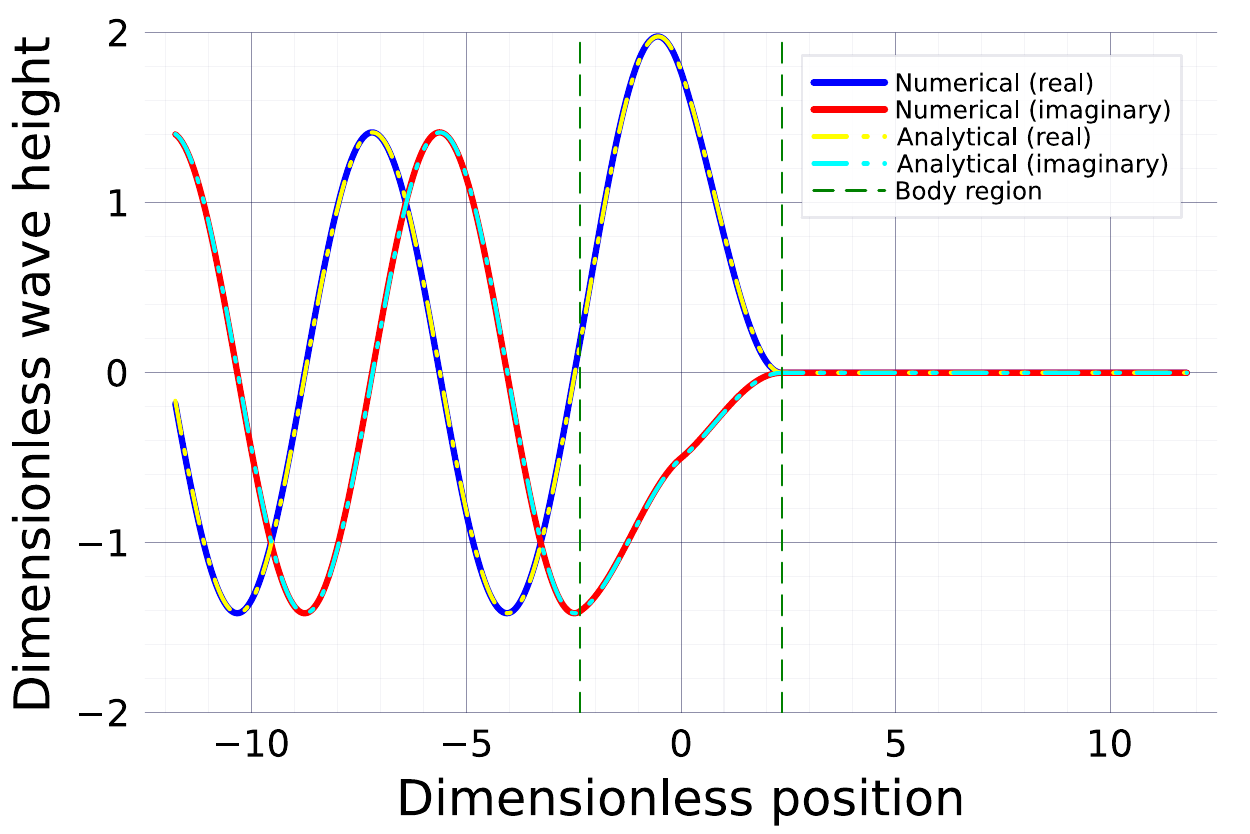}
        \caption{}
        \label{v0_pwr_h_plot}
    \end{subfigure}
    \caption{(a) Resulting plots for the real and imaginary parts of the source $\hat{Q}$ for the bounded power case where $\frac{U}{c}=v=0$ and $\frac{L\omega}{c}= l=3\pi/2$. (b) A similar plot for the wave-field $\hat{h}$ for the same case. There is no wave on the right implying this is the optimum result to start and travel rightward.}
    \label{fig:v0_pwr}
\end{figure}

To validate the numerical result with the analytical expression, we want to do something similar to what was done in Section \ref{subsec:v0_nrm}. This time, we have less information because the constraints do not give us information on $\hat{Q}$ like \eqref{eq:finding_DL_DR}. Therefore, as an ansatz we will prescribe $\hat{Q}$ to be a step function of the form
\begin{equation}\label{eq:step_defn}
    \hat{Q} = 
    \begin{cases}
        \alpha, & \textrm{if }\, x\in \left[-\frac{l}{2},0\right),
        \\
        \beta, & \textrm{if }\, x\in \left[0,\frac{l}{2}\right], 
        \\
        0, & \textrm{otherwise},
    \end{cases}
\end{equation}
and use the numerical optimisation code to find $\alpha, \beta \in \mathbb{C}$ which can then be used to validate $\hat{h}$. This will show that there is no wave on the right hand side as is the case in \eqref{eq:v0_amplitude_cond}. Given the step function in figure \ref{v0_pwr_q_plot}, the wave-field is found analytically and numerically and their comparison is demonstrated in figure \ref{v0_pwr_h_plot}. To fully validate the result, the condition for the the left-travelling wave in \eqref{eq:v0_amplitude_cond} and the thrust \eqref{eq:v0_thrust_expectation} are compared using relative error, which were within $0.008\%$ and $0.0069\%$ respectively. To validate the right hand wave, the relative error is undefined since the value we are comparing to is zero. We will instead present the absolute difference in the numerical value, which is $5.9\times10^{-8}$. Given the prescribed step function can be substituted into \eqref{eq:v0_amplitude_cond}, we can get constraints on $\alpha$ and $\beta$. However, we leave this calculation for Section \ref{subsec:sub_power} in the case of general subcritical $v$.

\section{Subcritical motion \texorpdfstring{$0<v<1$}{v<1}}\label{sec:subcritical}

Up to now, all that has been considered is the start-up case. 
This section will introduce a velocity $U$ as in figure \ref{fig:Setup}. 
In dimensionless form, the source is moving at a velocity $v = \frac{U}{c}$. 
The body velocity is scaled by the velocity of the waves emitted $c$ such that the Froude number 
can be interpreted like a Mach number, $\mathrm{Fr} = \frac{U}{\sqrt{gH}} = \frac{U}{c} = \mathrm{Ma}$. 
With this scaling, the dimensionless velocity is split into three different regimes: 
subcritical velocities ($0<v<1$) where the source is travelling slower than the waves 
it emits, supercritical velocities ($v>1$) where the source moves quicker than 
the waves it produces, and the critical velocity ($v=1$). We will discuss the latter 
two in Sections \ref{sec:supercritical} and \ref{sec:critical} respectively. 
To begin, we will rewrite the optimisation problem to include a subcritical velocity 
and discuss the two choices of bound similar to the previous section. 

As was the case in Section \ref{sec:startup}, the assumption of a periodic source is applied. Due to a moving body in a static frame, expressions for the wave-field and the source are 
\begin{align}\label{eq:sub_periodic_decomp}
    h(x,t) = \mathrm{Re}\left[\hat{h}(x-vt)e^{it}\right], & & Q(x,t) = \mathrm{Re}\left[\hat{Q}(x-vt)e^{it}\right].
\end{align}
A Galilean transformation is applied to work in the rest frame of the source. The Galilean transformation will define new spatial and temporal coordinates $\tilde{x}$ and $\tilde{t}$, 
\begin{align}\label{eq:gal_transform}
    \tilde{x} = x-vt & & \tilde{t} = t.
\end{align}
As such, the PDE in \eqref{eq:dimensionless_wave} is reduced to a modified ODE,
\begin{equation}\label{eq:main_ode}
    (1-v^2)\hat{h}''(\tilde{x}) + 2iv\hat{h}'(\tilde{x}) +\hat{h}(\tilde{x}) = -\hat{Q}(\tilde{x}).
\end{equation}
This is solved with the same Green's function approach as before:
\begin{equation}\label{eq:sub_ode_gen_sol}
    \hat{h}(\tilde{x}) = Ae^{\frac{i\tilde{x}}{1+v}} + Be^{-\frac{i\tilde{x}}{1-v}} + \frac{i}{2}\int^{\tilde{x}}_{-\frac{l}{2}}\hat{Q}(X)\left(e^{\frac{i\left(\tilde{x}-X\right)}{1+v}} - e^{\frac{i\left(X-\tilde{x}\right)}{1-v}}\right)\mathrm{d}X,
\end{equation}
where $A,B\in\mathbb{C}$ are constants to be found. Sommerfeld boundary conditions will be used again, but they now take a Doppler shifted form,
\begin{align}\label{eq:sub_radiative_bc}
    \hat{h}'\left(\pm\frac{l}{2}\right) = \mp ik_{\pm}\hat{h}\left(\pm\frac{l}{2} \right),
\end{align}
where $k_+ = \frac{1}{1-v}$, $k_- = \frac{1}{1+v}$ are the right- and left-travelling Doppler shifted wavenumbers respectively. The resulting wave field is given by 
\begin{equation}\label{eq:sub_h_sol}
    \hat{h}(\tilde{x}) = -\frac{i}{2}\int^{\frac{l}{2}}_{\tilde{x}}\hat{Q}(X)e^{\frac{i\left(\tilde{x}-X\right)}{1+v}}\,\mathrm{d}X - \frac{i}{2}\int^{\tilde{x}}_{-\frac{l}{2}}\hat{Q}(X)e^{\frac{i\left(X-\tilde{x}\right)}{1-v}}\,\mathrm{d}X.
\end{equation}

An equivalent expression for the thrust can be derived (see Appendix \ref{ap:thrust_derive}) using the same method as was done to get \eqref{eq:v0_L-H_result} and $\bar{F}_T$ will be as in \eqref{eq:v0_thrust_with_quarter}. In this case $k_{\pm}$ are the same as in \eqref{eq:sub_radiative_bc} and are interpreted as a correction due to the Doppler shift in the moving frame. Similarly, the time-averaged power in the moving case can be found. Following the steps taken in Appendix \ref{ap:power_derive}, the resulting expression for the time-averaged injected power is 
\begin{equation}\label{eq:vpower}
	\overline{\mathrm{Pow}} = \langle\hat{Q},i\hat{h}\rangle - v\langle \hat{Q},\hat{h}'\rangle,
\end{equation}
where the second term is proportional to the thrust as in \eqref{eq:v0_thrust_with_quarter}.

Given that we have the power and thrust, the efficiency $\eta$ can be discussed next. The efficiency will tell us how much of the injected power is used as thrust. In dimensionless form, all that is required is \eqref{eq:v0_thrust_with_quarter} and \eqref{eq:vpower} to provide this information
\begin{equation}
    \eta = \frac{\bar{F}_T}{\overline{\mathrm{Pow}}} = \frac{\langle \hat{Q},\hat{h}'\rangle}{\langle\hat{Q},i\hat{h}\rangle - v\langle\hat{Q},\hat{h}'\rangle}. 
\end{equation}
Note that in dimensional form, we see that the efficiency is given as $\eta = \bar{F}_T\cdot c/\overline{\mathrm{Pow}}$, because the energy is being radiated at speed $c$. In the start-up case, where the power was bounded to be $\overline{\mathrm{Pow}} = 1$ the resulting force was $\bar{F}_T=1$. Hence, the efficiency is $\eta=1$ when $\hat{Q}$ is optimised. If we had a purely symmetric source, the difference in fore-aft amplitude squared would be zero resulting in $\bar{F}_T=0$ and hence $\eta=0$. We will revisit the efficiency in the discussion of the results for the bounded norm and power in the subcritical case. 

\subsection{Case 1 - Bounded norm}

Similar to the start-up case (Section \ref{subsec:v0_nrm}), an analytical approach shows us that to be an optimal solution the source $\hat{Q}(\tilde{x})$ must satisfy
\begin{equation}
    \mathcal{L}_v\hat{Q} = 0,
\end{equation}
where $\mathcal{L}_v = (1-v^2)\partial_{\tilde{x}\tilde{x}} + 2iv\partial_{\tilde{x}} +\mathbb{I}$. This means that the source must be composed of a rightward and leftward Doppler-shifted wave 
\begin{equation}\label{eq:Q_wave_sols}
    \hat{Q}(\tilde{x}) = D_{\mathrm{L}}e^{\frac{i\tilde{x}}{1+v}}+D_{\mathrm{R}}e^{-\frac{i\tilde{x}}{1-v}},
\end{equation}
where conditions on $D_{\mathrm{L}},D_{\mathrm{R}}\in\mathbb{C}$ are found in Appendix \ref{norm_appendix}. These are summarised as follows:
\begin{equation}\label{eq:sub_norm_DL}
    |D_{\mathrm{L}}|^2 = 4\left(l+\frac{l(1-v^2)^2}{\left[2\lambda(1-v)-l\right]^2}\sin^2\left(\frac{l}{1-v^2}\right)+\frac{2(1-v^2)^2}{2\lambda(1-v)-l}\sin^2\left(\frac{l}{1-v^2}\right)\right)^{-1},
\end{equation}
\begin{equation}\label{eq:sub_norm_DR}
	|D_{\mathrm{R}}|^2 = \frac{|D_{\mathrm{L}}|^2(1-v^2)^2}{\big(2(1-v)\lambda-l\big)^2}\sin^2\left(\frac{l}{1-v^2}\right).
\end{equation}
Again, $\lambda\in\mathbb{R}$ is the Lagrange multiplier associated with the variational calculus, with eigenvalues
\begin{equation}
    	\lambda_{\pm} = \frac{vl \pm \sqrt{ l^2 - (1-v^2)^3\sin^2\left(\frac{l}{1-v^2}\right)}}{2(1-v^2)}.
\end{equation}
\begin{figure}
    \centering
    \begin{subfigure}[b]{0.45\textwidth}
        \centering
        \includegraphics[width=\textwidth]{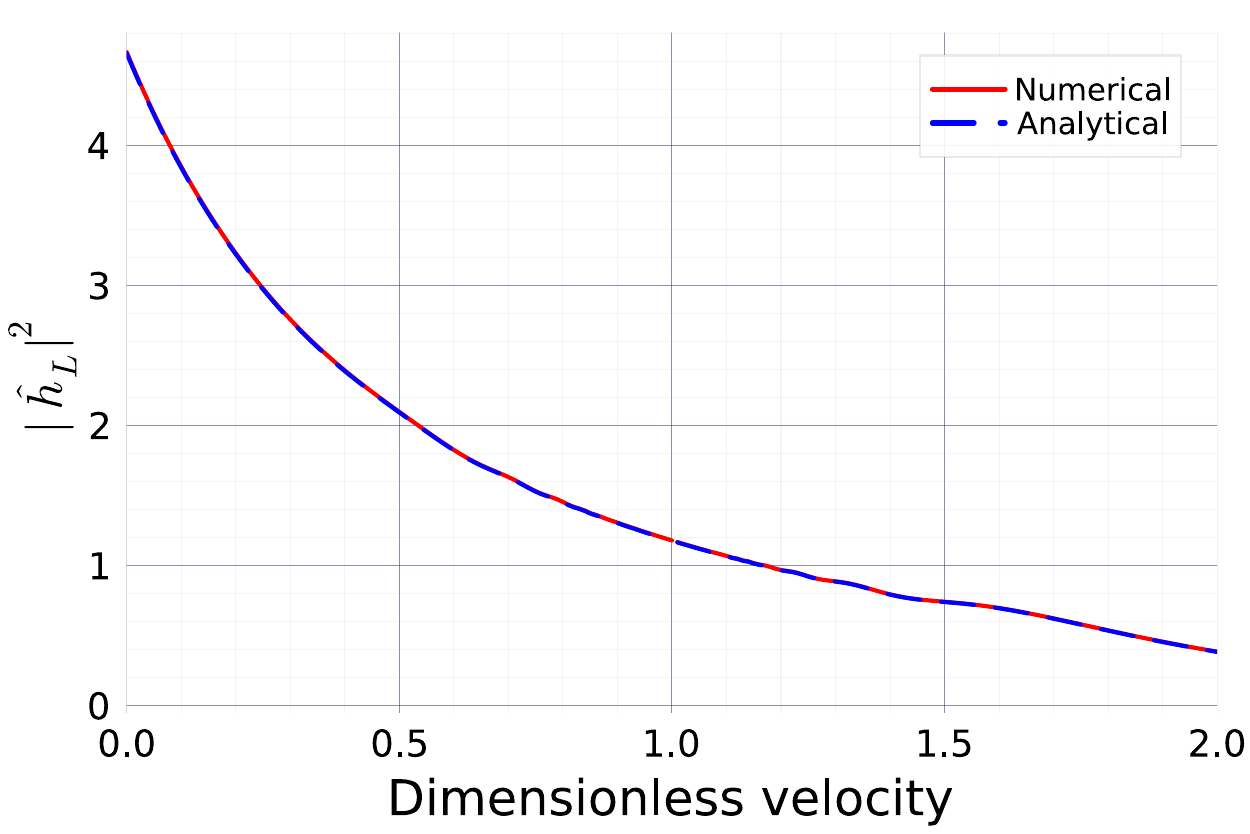}
        \caption{}
        \label{fig:norm_left}
    \end{subfigure}
    \begin{subfigure}[b]{0.45\textwidth}
        \centering
        \includegraphics[width=\textwidth]{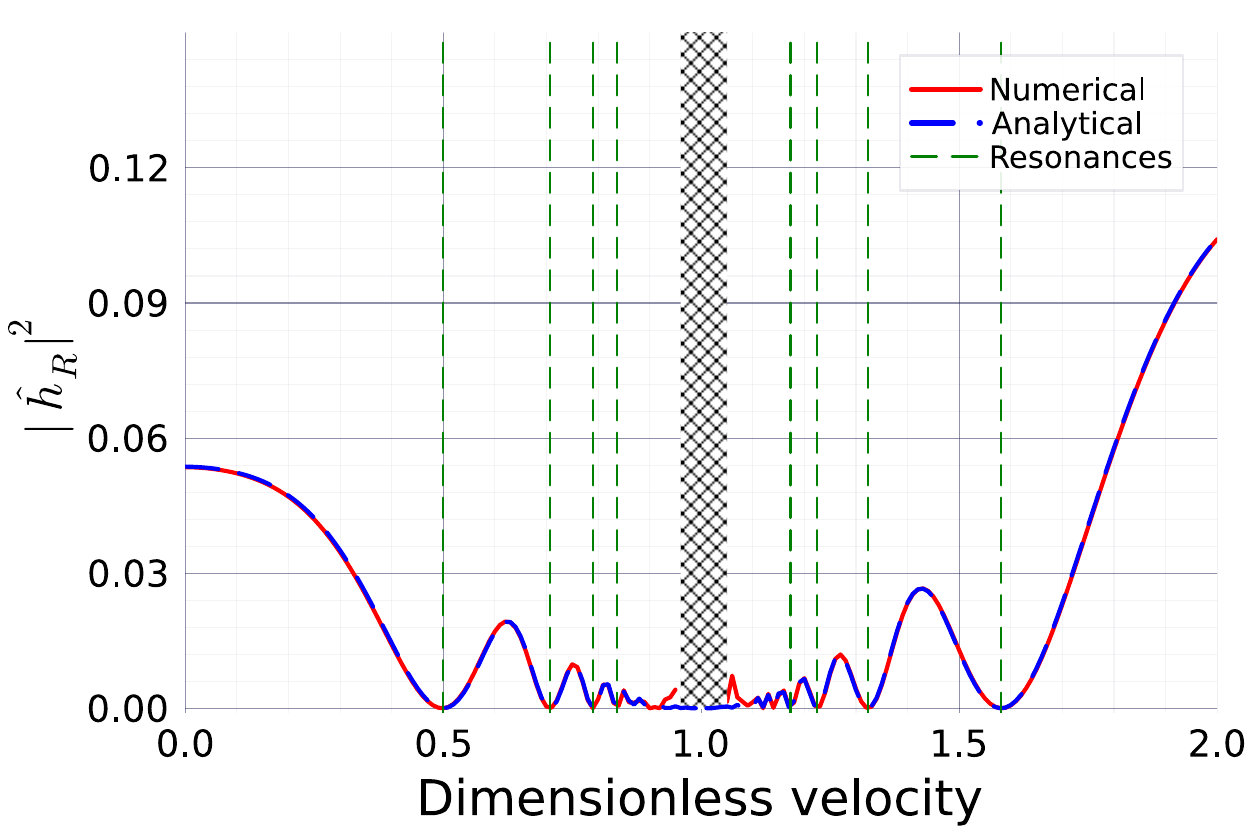}
        \caption{}
        \label{fig:norm_right}
    \end{subfigure}
    \caption{(a) Comparison between the analytically and numerically calculated 
    left and (b) right wave amplitudes squared for $l=3\pi/2$ over both subcritical 
    and supercritical velocities. For the right wave, the numerical scheme becomes 
    unstable close to $v=1$ and hence it has been hatched.}
    \label{fig:subcritical_norm_sweeps}
\end{figure}
It is worth noting than given $\lambda\in\mathbb{R}$, the constants $D_{\mathrm{L}}$ and $D_{\mathrm{R}}$ are in phase as they are related via multiplication by a real scalar (see \eqref{ap_eq:A-Brelation}). As was the case on start-up, the negative sign is taken to maximise the thrust for a body moving in the positive (right) direction. We validate that $\lambda\in\mathbb{R}$, since 
\begin{equation}\label{eq:sub_real_condition}
    l^2-(1-v^2)^3\sin^2\left(\frac{l}{1-v^2}\right)> 0,
\end{equation}
is satisfied for all real $l,v>0$. 

A study over the dimensionless velocity $v$ is demonstrated in figures \ref{fig:norm_left} and \ref{fig:norm_right} for $l=3\pi/2$. The forward wave appears to have zero amplitude at certain resonant values of $v$ and $l$, which is discussed later in Section \ref{sec:supercritical}. The expressions involving $D_{\mathrm{L}}$ and $D_{\mathrm{R}}$ can be compared with numerics. Using \eqref{ap_eq:DL_DR_to_numerics} to recover $\lambda^2|D_{\mathrm{L}}|^2$ and $\lambda^2|D_{\mathrm{R}}|^2$ in a similar way to \eqref{eq:finding_DL_DR}, we show correspondence within 0.221\% and 0.0005 respectively. The latter error is quoted as an absolute value since the right wave has zero amplitude at resonance points. 

\begin{figure}
    \centering
    \begin{subfigure}[b]{0.45\textwidth}
        \centering
        \includegraphics[width=\textwidth]{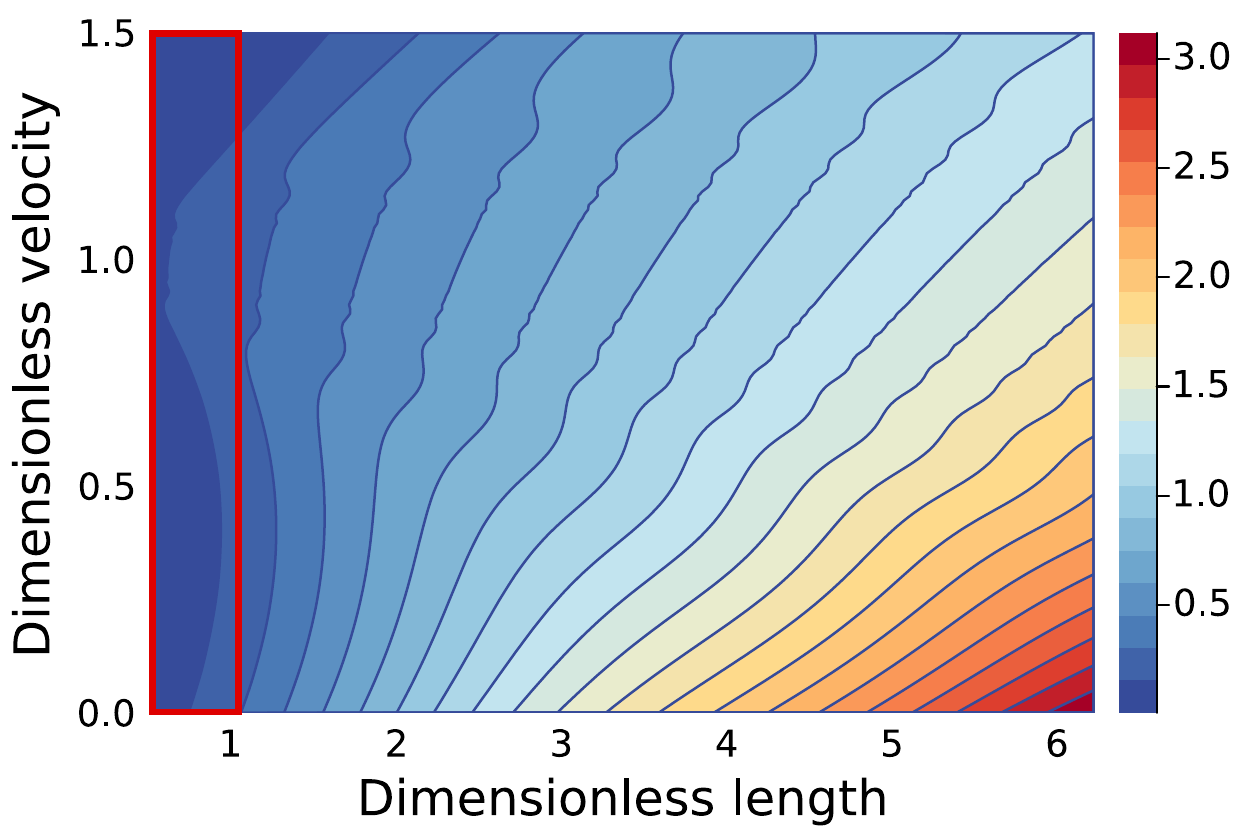}
        \caption{}
        \label{fig:heatmap_full}
    \end{subfigure}
    \begin{subfigure}[b]{0.45\textwidth}
        \centering
        \includegraphics[width=\textwidth]{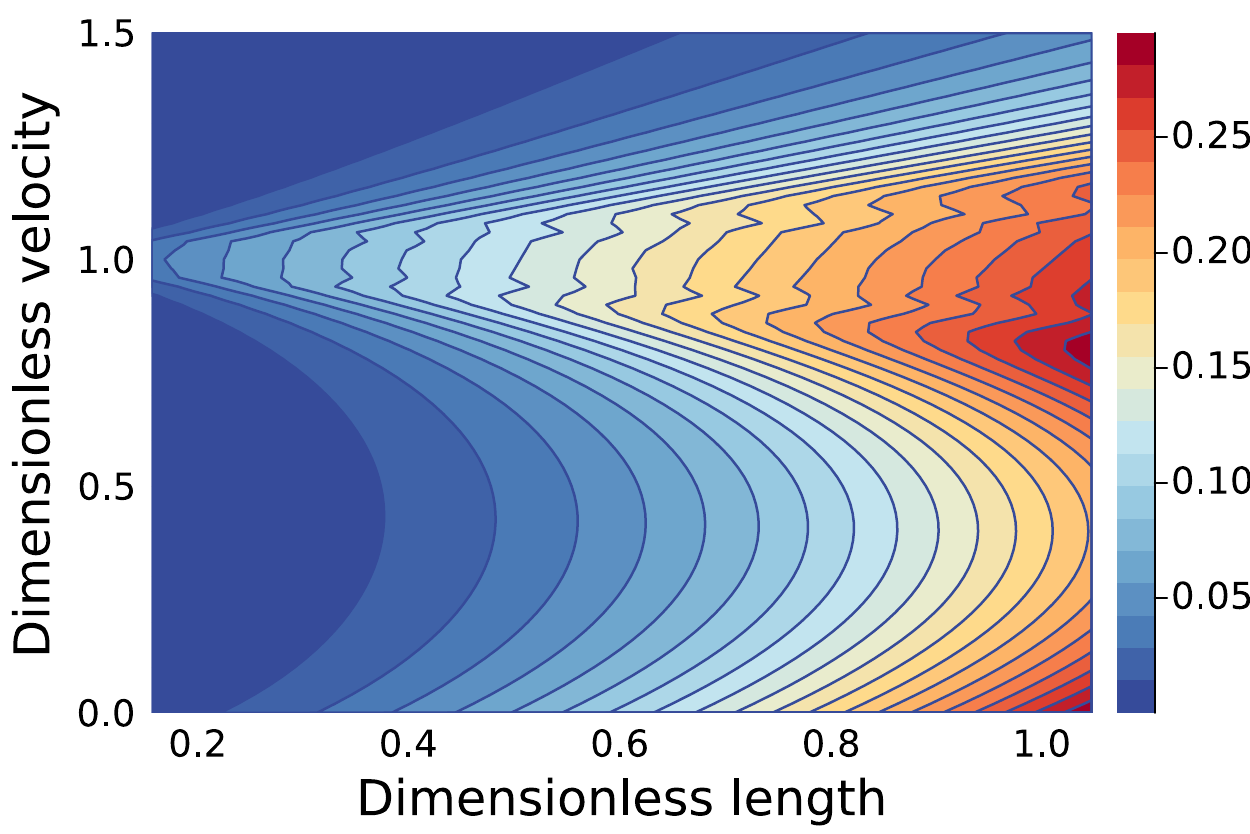}
        \caption{}
        \label{fig:heatmap_small}
    \end{subfigure}
    \caption{(a) Contour plot where colour corresponds to the magnitude of the force. For a given $v$, a larger time-averaged thrust can be achieved with a larger $l$. (b) The same heat map over a range of smaller $l$ denoted by the red box in (a) to convey the trend for smaller length scales, demonstrating maximal thrust at dimensionless velocities closer to 1.}
    \label{fig:heatmaps}
\end{figure}

Since there is no forward wave in some conditions, we ask, is there an optimal $l$ for a given $v$? The contour plots in figure \ref{fig:heatmaps} best illuminate that the same trend as in figure \ref{fig:v0_lsweep} continues for subcritical values of $v$. 
\begin{figure}
    \centering
    \begin{subfigure}[b]{0.45\textwidth}
        \centering
        \includegraphics[width=\textwidth]{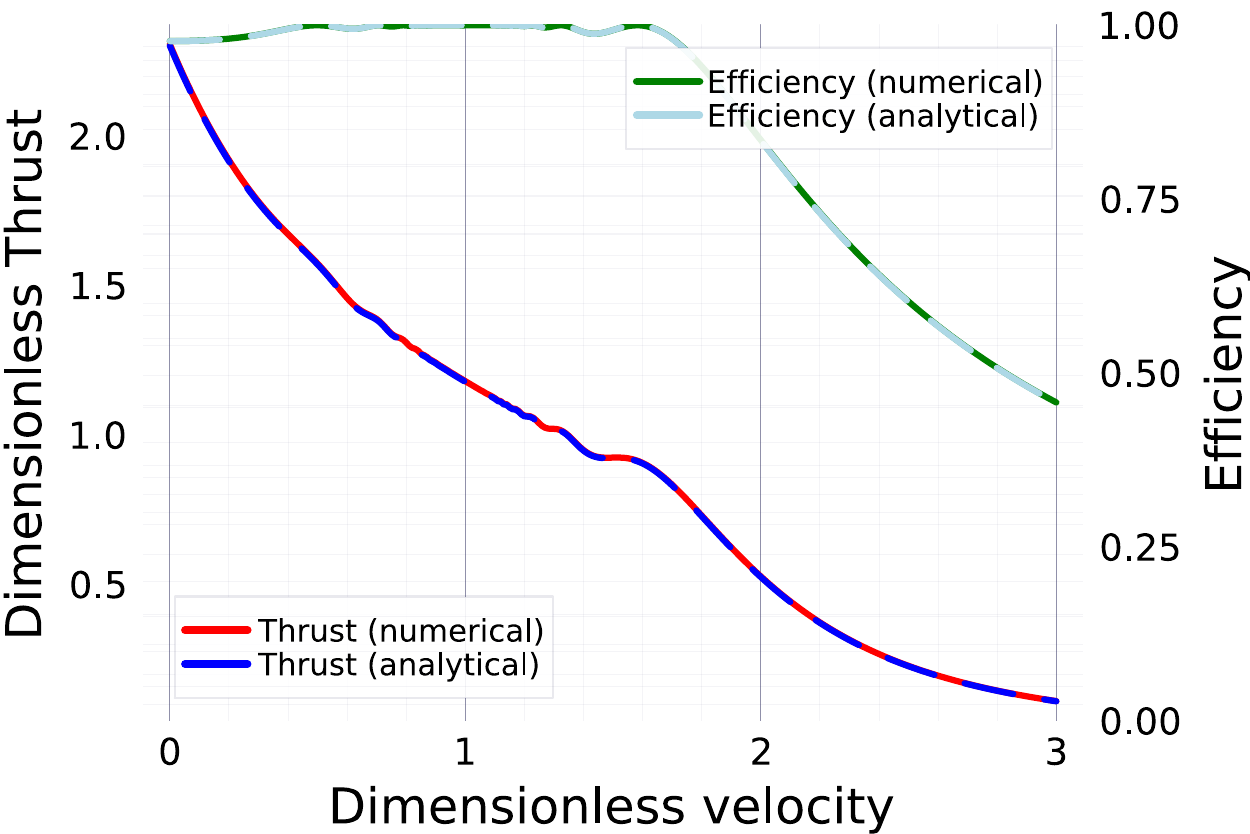}
        \caption{}
        \label{fig:norm_force_pi}
    \end{subfigure}
    \begin{subfigure}[b]{0.45\textwidth}
        \centering
        \includegraphics[width=\textwidth]{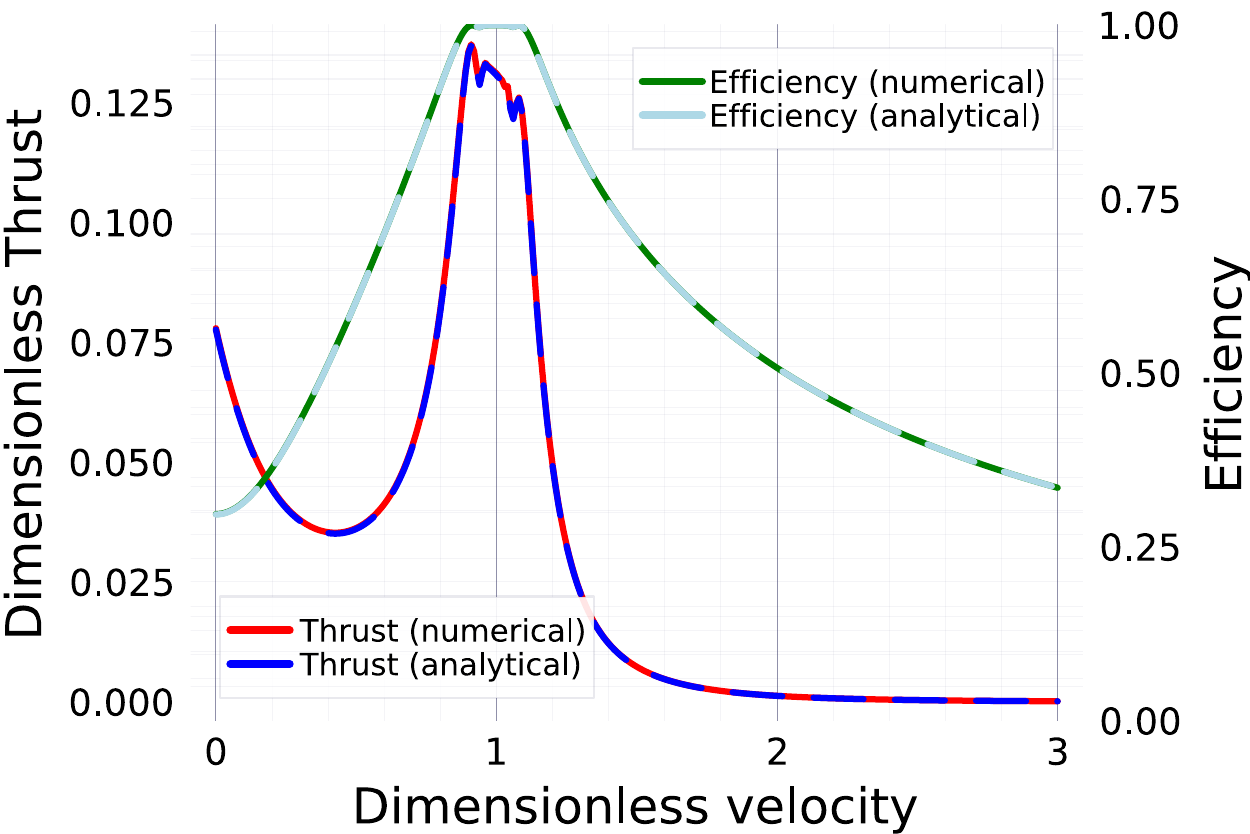}
        \caption{}
        \label{fig:norm_force_pi/6}
    \end{subfigure}
    \caption{(a) Time-averaged thrust over velocities from $v=0$ to $v=3$ for $l=3\pi/2$. The efficiency is also plotted on the right axis showing the reduction after the last resonance point. (b) A similar plot for $l=\pi/6$ demonstrates that the region of maximal efficiency reduces due to the last and first resonance points being closer to $v=1$ for smaller values of $l$.} 
    \label{fig:supercritical_norm_forces}
\end{figure}
For a given $v$, maximal thrust is achieved at maximal $l$. However, there is an interesting behavioural difference between smaller and larger length scales. For a fixed $l$ that is large, the force is maximal close to $v=0$, but as we shrink the length scale, the optimum begins to move towards $v=1$, which is demonstrated in figure \ref{fig:heatmap_small}. For a given short $l$, the time-averaged thrust may be non-monotone with velocity, which is demonstrated in the subcritical part of figure \ref{fig:norm_force_pi/6}.  The efficiency is also plotted showing that $\eta=1$ is achieved for a small set of velocities close to $v=1$ for small $l$. 
This will be shown in greater detail in Section \ref{sec:supercritical}. 

\subsection{Case 2 - Bounded power}\label{subsec:sub_power}

The general approach taken while moving is the same as it is in the start-up case. The variational calculus approach (Appendix \ref{power_appendix}) produces two results depending on the choice of the solution for the Lagrange multiplier, $\lambda$:
\begin{align}
        C_{\mathrm{L}}\,e^{\frac{i\tilde{x}}{1+v}} + C_{\mathrm{R}}\,e^{-\frac{i\tilde{x}}{1-v}} = C\,e^{\frac{i\tilde{x}}{1+v}} & & & \textrm{for  } \lambda_1 = -1,
        \label{eq:sub_cond_in_use}
        \\ 
        C_{\mathrm{L}}\,e^{\frac{i\tilde{x}}{1+v}} + C_{\mathrm{R}}\,e^{-\frac{i\tilde{x}}{1-v}} = C\,e^{-\frac{i\tilde{x}}{1-v}} & & & \textrm{for}\,\, \lambda_2 = 1 ,
\end{align}
where 
\begin{align}\label{sub_equated_coeffs}
    C_{\mathrm{L}} = -\frac{i}{2}\int^{\frac{l}{2}}_{-\frac{l}{2}}\hat{Q}(X)e^{-\frac{iX}{1+v}}\,\textrm{d}X, & & C_{\mathrm{R}} = -\frac{i}{2}\int^{\frac{l}{2}}_{-\frac{l}{2}}\hat{Q}(X)e^{\frac{iX}{1-v}}\,\textrm{d}X.
\end{align}
The choice of $\lambda$ governs whether we are going to maximise the thrust in the positive or negative direction. Given that the subcritical velocity is positive ($0<v<1$), to maximise thrust in the positive direction, $\lambda_1$ is chosen. Using \eqref{eq:sub_cond_in_use}, $C_{\mathrm{L}}=C$ and $C_{\mathrm{R}}=0$. The power constraint \eqref{eq:vpower} is then used to give
\begin{align}\label{eq:sub_c1_c2_cond}
    \left|-\frac{i}{2}\int^{\frac{l}{2}}_{-\frac{l}{2}}\hat{Q}(X)e^{-\frac{iX}{1+v}}\,\textrm{d}X\right|^2 = 2(1+v), && \left|-\frac{i}{2}\int^{\frac{l}{2}}_{-\frac{l}{2}}\hat{Q}(X)e^{\frac{iX}{1-v}}\,\textrm{d}X\right|^2 = 0
\end{align}
and this can be used to show that for all subcritical $v$,
\begin{equation}\label{eq:sub_optimum_thrust}
    \bar{F}_T = \frac{1}{4}\left(\frac{2}{1+v}\left|C_{\mathrm{L}}\right|^2 - \frac{2}{1-v}\left|C_{\mathrm{R}}\right|^2\right) = 1.
\end{equation}
The above results can be verified numerically over the interval $0\leq v<1$. The condition for the left travelling wave \eqref{eq:sub_c1_c2_cond} was tested and the infinity norm of the relative error at each point sampled was taken to show that all the points were within $0.0010\%$. Through a similar test the force \eqref{eq:sub_optimum_thrust} was shown to be within $0.0011\%$. As discussed in Section \ref{subsec:v0_power}, the right hand wave condition of zero \eqref{eq:sub_c1_c2_cond} does not permit the calculation of the relative error so we will just provide the infinity norm of the absolute difference between the analytical and numerical solution, which is $1.11\times10^{-5}$. 

This is a very general solution since there are infinite solutions for $\hat{Q}$ that satisfy \eqref{eq:sub_c1_c2_cond}. It is also interesting that we are free to choose any $l$ and still maintain an optimal solution that has $\bar{F}_T = 1$. This corresponds to an efficiency of $\eta=1$ since the power and force are found to be equal. 

To shed some light onto what an example solution would look like, we will return to the example used in Section \ref{subsec:v0_power} where $\hat{Q}(\tilde{x})$ is defined as a step function in the non-zero domain \eqref{eq:step_defn}. 
The conditions \eqref{eq:sub_c1_c2_cond} become 
\begin{gather}
    \left|-\frac{i}{2}\left(\alpha\int^{0}_{-\frac{l}{2}} e^{-\frac{iX}{1+v}}\,\textrm{d}X+\beta\int^{\frac{l}{2}}_{0} e^{-\frac{iX}{1+v}}\,\textrm{d}X\right)\right|^2 = 2(1+v),
    \label{eq:subcrit_step_cond1}
    \\
    \left|-\frac{i}{2}\left(\alpha\int^{0}_{-\frac{l}{2}} e^{\frac{iX}{1-v}}\,\textrm{d}X + \beta\int^{\frac{l}{2}}_{0} e^{\frac{iX}{1-v}}\,\textrm{d}X\right)\right|^2=0.
    \label{eq:subcrit_step_cond2}
\end{gather}
Equation \eqref{eq:subcrit_step_cond2} gives us 
\begin{equation}\label{eq:sub_pwr_alpha_beta_relation}
    \alpha = \beta \left(\frac{1-e^{\frac{il}{2(1-v)}}}{1-e^{-\frac{il}{2(1-v)}}} \right) = \beta \,\zeta,
\end{equation}
where $\zeta\in\mathbb{C}$. This can be substituted back into \eqref{eq:subcrit_step_cond1} to show that 
\begin{equation}\label{eq:sub_pwr_beta_square}
    |\beta|^2 = 4\left(\left|\zeta\left(1-e^{\frac{il}{2(1+v)}}\right) + e^{-\frac{il}{2(1+v)}} -1\right|^2\right)^{-1}.
\end{equation}
If we wish to prescribe that $\hat{Q}$ be a step function of this form, the constants $\alpha$ and $\beta$ are constrained to satisfy the condition on their modulus squared, and are only defined up to an arbitrary phase, similar to the bounded norm case. In contrast to the bounded norm case, there is a phase difference between $\alpha$ and $\beta$ since $\zeta \in \mathbb{C}$.

\section{The critical case (\texorpdfstring{$v=1$}{v=1})}\label{sec:critical}

In order to study the whole range of velocities, the critical velocity case ($v=1$) must be studied separately. This is a simple case because \eqref{eq:main_ode} reduces to a first order ODE,
\begin{equation}\label{eq:critical_ode}
        2i\hat{h}'(\tilde{x}) +\hat{h}(\tilde{x}) = -\hat{Q}(\tilde{x}).
\end{equation}
The sole boundary condition required is that the fluid to the right of the body is unperturbed, since the body is now moving as quickly as the waves that it emits. The boundary condition takes the form
\begin{align}
    \hat{h}= 0: \,\,\,\tilde{x} > \frac{l}{2}.
\end{align}
The resulting particular solution for the wave-field $\hat{h}(\tilde{x})$ is
\begin{equation}
    \hat{h}(\tilde{x}) = -\frac{i}{2}\int^{\frac{l}{2}}_{\tilde{x}} \hat{Q}(X)e^{\frac{i(\tilde{x}-X)}{1+v}}\,\mathrm{d}X,    
\end{equation}
where $v=1$. Given this expression for $\hat{h}$, we can recover the thrust through radiation stresses 
\begin{equation}
    \frac{1}{2}k_{-}|\hat{h}_{-}|^2 = \langle\hat{Q},\hat{h}'\rangle,
\end{equation}
where $k_{-}=\frac{1}{1+v}$ is the wavenumber for the left travelling wave with amplitude $|\hat{h}_{-}|$. This is the same form as \eqref{eq:v0_L-H_result} but the fore wave is zero.

The same analytical and numerical methods are used as the subcritical case and we will briefly outline the results. In the bounded norm case there will only be a wave in the aft direction which from the variational calculus is generated by $\hat{Q}$ such that,
\begin{equation}
    \hat{Q}(\tilde{x}) = D_{\mathrm{L}}e^{\frac{i\tilde{x}}{2}},
\end{equation}
where $D_{\mathrm{L}}$ is found to satisfy 
\begin{equation}\label{eq:crit_norm_result}
    |D_{\mathrm{L}}|^2 = \frac{4}{l}.
\end{equation}
Since there is no forward wave, efficiency of $\eta=1$ is achieved in this case. 
The Lagrange multiplier is found to be $\lambda = -\frac{l}{4}$ and the condition in \eqref{eq:crit_norm_result} can be tested using expressions found in a similar way to \eqref{ap_eq:DL_DR_to_numerics} and $\lambda^2|D_{\mathrm{L}}|^2$ is found to have a relative error of $0.0002\%$.

For the bounded power case, the condition is the first equation in \eqref{eq:sub_c1_c2_cond} when $v=1$ (the second condition is automatically satisfied due to the impossibility of a wave on the right hand side). Hence our condition is 
\begin{equation}\label{eq:crit_v1_condition}
    \Bigg|-\frac{i}{2}\int^{\frac{l}{2}}_{-\frac{l}{2}} \hat{Q}(X)e^{-\frac{iX}{2}}\,\mathrm{d}X\Bigg|^2 = 4.
\end{equation}
The resulting force is again $\bar{F}_T =1$. These two results are validated with the numerical approach with relative errors of $0.037\%$ and $6.07\times10^{-12}\%$ respectively.

\section{Supercritical motion \texorpdfstring{$v>1$}{v>1}}\label{sec:supercritical}

The final part of our study is the supercritical velocity regime. If we recall, the dimensionless velocity, $v=U/c$, is the ratio of the body velocity and the emitted wave velocity. The $v>1$ case, corresponds to the body moving faster than the waves that it emits. Hence, any wave emitted will immediately be overtaken by the body. Similar to the subcritical case, a Galilean transformation into the moving frame is done using the transformation in \eqref{eq:gal_transform}. 

The radiative boundary conditions applied in Sections \ref{sec:startup} and \ref{sec:subcritical} are no longer suitable because no wave can possibly propagate on the right hand side and the left hand side will be composed of two waves of differing wavenumbers. We can use the fact that there is no wave on the right to our advantage by applying the following boundary conditions 
\begin{align}\label{eq:supercritical_bcs}
    \hat{h} = \hat{h}' = 0: \,\,\, \tilde{x} > \frac{l}{2}.
\end{align}
The Green's function approach to solving \eqref{eq:main_ode} is applied again to calculate the solution for the wave-field, $\hat{h}(\tilde{x})$:
\begin{equation}\label{eq:supercritical_h}
    \hat{h}(\tilde{x}) = -\frac{i}{2}\int^{\frac{l}{2}}_{\tilde{x}} \hat{Q}(X)e^{\frac{i(\tilde{x}-X)}{1+v}}\,\textrm{d}X +\frac{i}{2}\int^{\frac{l}{2}}_{\tilde{x}} \hat{Q}(X)e^{\frac{i(\tilde{x}-X)}{v-1}}\,\textrm{d}X,
\end{equation}
where we have swapped $-(1-v)$ with $v-1$ to emphasise that this solution is now composed of two left travelling waves in the rest frame of the body.

A similar approach of recovering the radiation stress is applied to get the expression for the thrust as in Appendix \ref{ap:thrust_derive}. The following equivalence relation is recovered:
\begin{equation}\label{eq:sup_L-H_result}
    -\frac{1}{2}\left[k|\hat{h}|^2\right]^{\rightarrow}_{\leftarrow} = \langle\hat{Q},\hat{h}'\rangle,
\end{equation}
where $\rightarrow$ and $\leftarrow$ denote the right and left travelling waves in the rest frame respectively. This choice of notation  represents a subtlety of the supercritical case where both waves are left-travelling waves in the moving frame. But once we translate back into the rest frame, there is a forward and backwards wave. As such, $k$ represents the Doppler shifted wavenumber in the moving frame where $k_{\leftarrow}=\frac{1}{1+v}$ and $k_{\rightarrow} = \frac{1}{v-1}$.

Variational calculus is undertaken in a similar way to the subcritical case. The bounded norm and power cases will be studied again, but since there are only minor changes, rather than being split into sections, we simply summarise the results.

In the bounded norm case the optimal condition again is that $\hat{Q}(\tilde{x})$ is composed of wave-like parts as was the case in \eqref{eq:Q_wave_sols}. This time the corresponding $|D_{\mathrm{L}}|^2$, $|D_{\mathrm{R}}|^2$ and $\lambda$ are 
\begin{equation}\label{eq:sup_norm_DL}
    |D_{\mathrm{L}}|^2 = 4\left(l+\frac{l(v^2-1)^2}{\left[2\lambda(v-1)-l\right]^2}\sin^2\left(\frac{l}{v^2-1}\right)+\frac{2(v^2-1)^2}{2\lambda(v-1)-l}\sin^2\left(\frac{l}{v^2-1}\right)\right)^{-1},
\end{equation}
\begin{equation}\label{eq:sup_norm_DR}
	|D_{\mathrm{R}}|^2 = \frac{|D_{\mathrm{L}}|^2(v^2-1)^2}{\big(2(v-1)\lambda-l\big)^2}\sin^2\left(\frac{l}{v^2-1}\right),
\end{equation}
\begin{equation}\label{eq:sup_norm_lambda}
    \lambda_{\pm} = \frac{l \pm \sqrt{v^2l^2 - (v^2-1)^3\sin^2\left(\frac{l}{v^2-1}\right)}}{2(v^2-1)}.
\end{equation}
Similar to before, to maximise the force in the positive direction, the negative sign is taken in the expression for $\lambda\in\mathbb{R}$. The analytical and numerical results are verified as they were in the subcritical case. Looking again at the supercritical part of figure \ref{fig:norm_right}, the forward wave demonstrates a resonance behaviour where at certain velocities it has zero amplitude. Turning to the analytical expressions above, we can see that the $\sin^2(l/(v^2-1))$ term in \eqref{eq:sup_norm_DR} is important. To verify this, we look at the cases where $|D_{\textrm{R}}|^2=0$. To also link to the subcritical case we will use the fact that $\sin^2(l/(v^2-1)) = \sin^2(l/(1-v^2))$ and find
\begin{align}\label{eq:resonance_velocities}
    \sin^2\left(\frac{l}{1-v^2}\right) = 0 & \implies   v = \left(1-\frac{l}{n\pi}\right)^{\frac{1}{2}},
\end{align}
where $n\in\mathbb{Z}\backslash \{0\}$. The positive square root is taken on the right hand side since $v\geq0$. In this current form, we see that the subcritical resonances will be seen for all possible $0<n<\frac{l}{\pi}$ while supercritical resonances for all $n<0$. Equation \eqref{eq:resonance_velocities} is used to find the velocities where the right hand wave is zero and is validated in figure \ref{fig:norm_right}. For the supercritical resonances, the largest resonant velocity will be found when $n=-1$. 

We have shown that for a given $l$, we can find values of $v$ for which the front wave will be zero. Inspired by the bounded power results, we can use this to our advantage in the interest of efficiency. For supercritical velocities this will be
\begin{equation}\label{eq:sup_nrm_eff}
    \eta = \frac{\bar{F}_T}{\overline{\mathrm{Pow}}} = \frac{-2(v-1)\lambda^2|D_{\textrm{R}}|^2 + 2(1+v)\lambda^{2}|D_{\textrm{L}}|^2}{2(v-1)\lambda^2|D_{\textrm{R}}|^2 + 2(1+v)\lambda^{2}|D_{\textrm{L}}|^2}.
\end{equation}
It is clear from \eqref{eq:sup_nrm_eff} that $|D_{\textrm{R}}|=0$ will produce maximal efficiency (see figure \ref{fig:norm_force_pi} at the values of $v$ where $\eta=1$) which is achieved by wasting no power on the wave that inhibits thrust. 

We note that $\eta$ drops for larger $v$ because the body is too short to generate a substantial wave-field to generate thrust. This is demonstrated in figure \ref{fig:norm_force_pi/6} where the range of values that efficiency is maximal appears to narrow down for small $l$. This manifests itself in \eqref{eq:resonance_velocities}. The last velocity value before efficiency diminishes at the $n=-1$ resonance will reduce as $l$ reduces. 

In contrast, the bounded power case will continue to follow the same result in \eqref{eq:sub_c1_c2_cond} where the only change is a $\frac{i}{2}$ rather than $-\frac{i}{2}$ preceding the second integral. Since the absolute value is taken, this has no effect. After inserting these optimal conditions into the expressions for the force, we find that $\bar{F}_T=1$ still holds. Hence the bounded power case results in $\eta=1$ for all $v>0$. The ansatz \eqref{eq:step_defn} can be used again here but we do not include the results since they are similar. 

\section{Quasi-periodic theory for a changing \texorpdfstring{$v$}{v}}\label{sec:quasiperiodic}

We have studied the optimal solution for a given velocity and found that in the bounded power case the time-averaged thrust is a constant ($\bar{F}_T=1$). All this work has been done in the absence of acceleration. We can now ask if we could create a model that introduces a small acceleration to move between different velocities while maintaining the optimal solution and the assumption that the waves are periodic? A dimensional acceleration such that $\dot{U}\ll\omega c$, permits the use of the previous results where acceleration can be interpreted as changing from one moving frame to the next. 

In order to make sense of moving between different velocities, the cruising velocity $v_\star$ must be known. For a given set of conditions, the balance between thrust $F_T$ and drag $F_D$ will return the equilibrium velocity. Given the expression for the time-averaged dimensionless drag force, the result from bounding the power will be used to create the aforementioned force balance and find the intercept which will be referred to as the cruising velocity. The form of $v_\star$ can then be used to understand how slowly varying the parameters of the problem can be used to move from $v=0$ to $v>1$. 

To find $v_\star$, we need to introduce a drag force. It is important to note that the true drag may contain contributions from wave drag, form drag due to shape, and shallow water resonance,
but we skip these details here since they aren't the focus of the study. Instead, we use a simple drag law to demonstrate how the force changes with speed. Therefore the dimensional drag will be considered to be,
\begin{equation}\label{eq:dimensional_drag}
    F_{D} = \frac{1}{2}C_{D} \rho U^2 L,
\end{equation}
where $C_D$ is a general drag coefficient. Similar to \citet{Benham2024propulsion}, we will neglect oscillatory drag and focus solely on horizontal drag.
To get the drag in dimensionless form, the dimensional drag is set equal to the dimensional thrust, 
\begin{equation}\label{eq:dimensional_thrust}
	\bar{F}_T = \rho g H^2 \langle\hat{Q},\hat{h}'\rangle.
\end{equation}
Hence, by dividing by $\rho g H^2$, the corresponding time-averaged dimensionless drag is found to be
\begin{equation}\label{eq:drag}
	\bar{F}_D = \frac{1}{2}C_Dv^2\frac{L}{H}. 
\end{equation}
In the previous sections we chose $\delta=1$ for the bounded power case to compare to the numerical work. In this section we will return to the general result $\overline{\textrm{Pow}}\leq\delta$. The resulting optimal time-averaged thrust is $\bar{F}_T = \delta$ which gives us the cruising velocity $v_\star$,
\begin{equation}\label{eq:force_drag_qperiodic}
\begin{aligned}
	   \delta = \frac{1}{2}C_D v^2\frac{L}{H} && \implies && 	v_{\star}(\delta) = \sqrt{\frac{2H\delta}{LC_D}}.
\end{aligned}
\end{equation}
The drag coefficient $C_D$, the body length $L$ and the fluid height $H$ are fixed which leaves the power injected $\delta$ as the only parameter to control the velocity. The velocity and power injected are related by $v_\star \sim \delta^{\frac{1}{2}}$. Using the thrust result $\bar{F}_T=\delta$, the optimal conditions on the source $\hat{Q}$ can be shown to also increase with $\delta$ (see \eqref{eq:quasiperiodic_conditions} below). In terms of $\delta$ and $v_{\star}$, the variational calculus produces the following conditions,
\begin{align}\label{eq:quasiperiodic_conditions}
    \left|-\frac{i}{2}\int^{\frac{l}{2}}_{-\frac{l}{2}}\hat{Q}(X)e^{-\frac{iX}{1+v_{\star}}}\,\mathrm{d}X\right|^2 = 2\delta(1+v_{\star}(\delta)), & & \left|-\frac{i}{2}\int^{\frac{l}{2}}_{-\frac{l}{2}}\hat{Q}(X)e^{\frac{iX}{1-v_{\star}}}\,\mathrm{d}X\right|^2=0.
\end{align}
Hence, as the velocity increases, we need a larger wave to maintain the optimal thrust. 

Given a small acceleration we can vary the cruising velocity $v_{\star}$ with the time-averaged power injected $\delta$. From here we can make two observations. To move quicker, one has to provide more power and to remain moving efficiently, the aft wave gets larger while the fore wave remains zero. This makes physical sense because the faster velocity requires that we generate more thrust which requires a bigger difference in fore-aft amplitude as is the case in \eqref{eq:v0_L-H_result}. Modifications can be made to this to delve further into the conditions for accelerating from start-up. Motivated by previous sections, we could prescribe $\hat{Q}$. For example, if we used a step function like \eqref{eq:step_defn}, we could study how the conditions for $\alpha$ and $\beta$ evolve in terms of the changing speed $v_{\star}$. We could also study adding in another constraint on top of the bounded power. If the norm is bounded simultaneously, there will be an upper bound on the magnitude of the source and hence there will come a point where the thrust will diminish from achieving $\bar{F}_T = \delta$ and efficiency will suffer. Since this section is a demonstration of a simple theory, these modifications will not be investigated here. 

\begin{figure}
    \centering
        \begin{subfigure}[b]{0.45\textwidth}
        \centering
        \includegraphics[width=\textwidth]{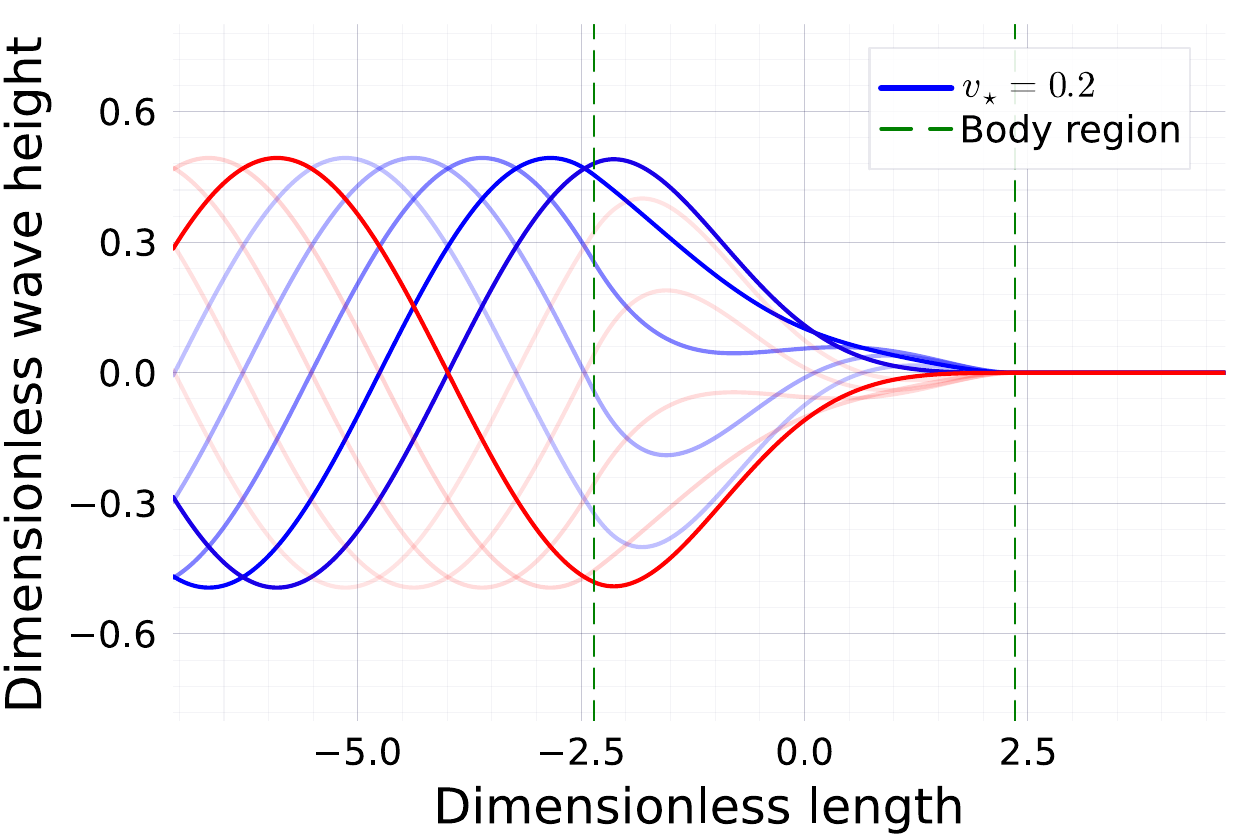}
        \caption{}
        \label{fig:t_evol_single}
    \end{subfigure}
        \begin{subfigure}[b]{0.45\textwidth}
        \centering
        \includegraphics[width=\textwidth]{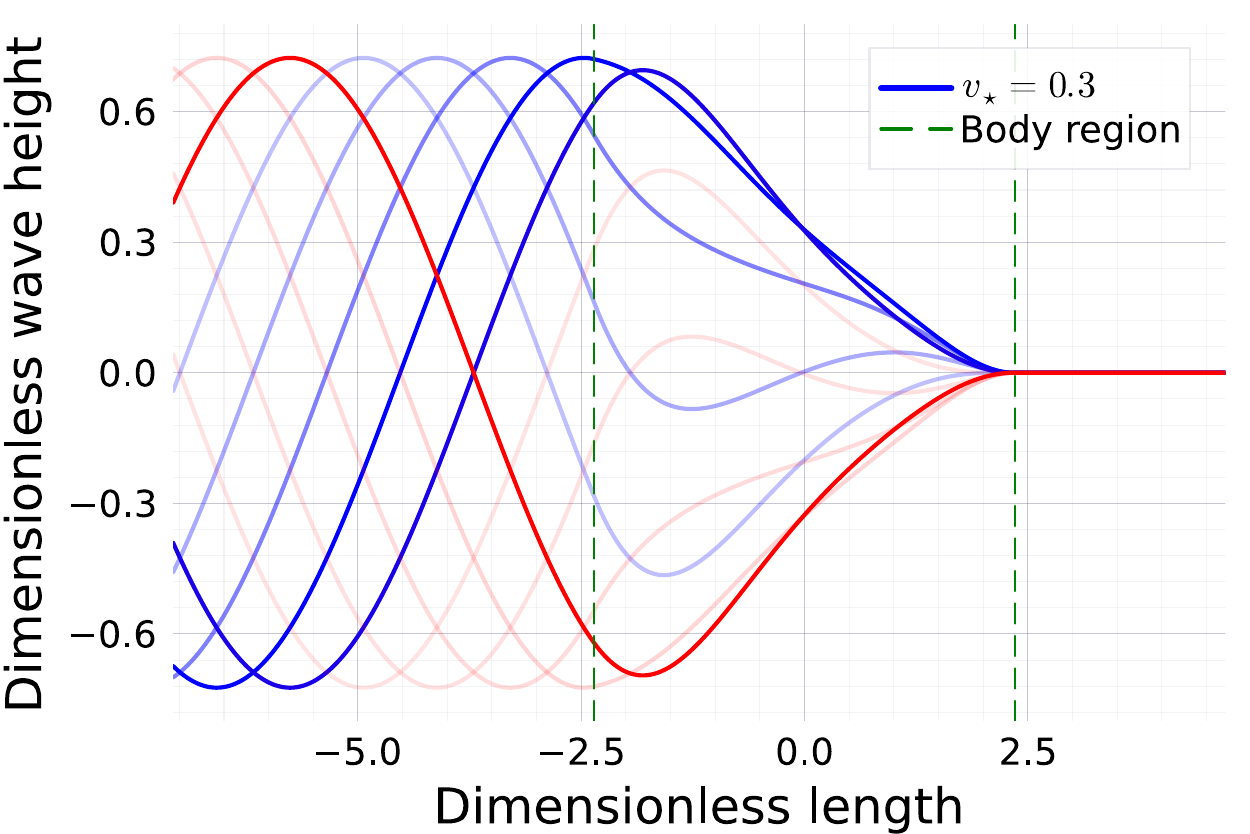}
        \caption{}
        \label{fig:t_evol_2}
    \end{subfigure}
    \caption{(a) Demonstration of the one period of oscillation where $v_{\star}=0.2$ and $l=3\pi/2$. The body is defined as the region between the vertical lines ($\tilde{x}\in\left[-l/2,l/2\right]$). The period is divided into $10$ points where the first half period is in blue and the latter in red. (b) A similar plot showing the time evolution at $v_{\star}=0.3$ demonstrating that the amplitude must increase to maintain the optimal solution at the larger velocity.}
    \label{fig:t-evol}
\end{figure}

To conclude this section, we will take this opportunity to look at the time-evolution of the solutions. We will reintroduce the temporal part and study the wave-field $h(x,t)$ as it is in \eqref{eq:sub_periodic_decomp} to produce a graph showing one whole period of oscillation. The numerically optimised bounded power solution is plotted over one period in figure \ref{fig:t_evol_single} (note, this is just one optimal $\hat{Q}$ out of infinitely many that satisfy \eqref{eq:sub_c1_c2_cond}). To visualise the results of \eqref{eq:quasiperiodic_conditions}, a comparison between a faster cruising velocity is shown in figure \ref{fig:t_evol_2}. Here we can see that the amplitude of oscillation is required to be greater to maintain optimal efficiency at the larger velocity.

\section{Discussion}\label{sec:Discussion}

In this paper, we have used gravity waves to generate a thrust to optimally push a body along the shallow water surface. Given one of two constraints where a bound is placed on either the power or the norm of the source, the conditions for optimal thrust are found. The analytical work sheds light on the choice of solution to optimise thrust while the numerical work is a validation step through solving the problem in a different way. A dimensionless velocity based on the Froude number ($v=U/c=U/\sqrt{gH}$) is used to divide the work into subcritical, critical and supercritical velocities. Studying each of these velocities using a Galilean transformation to operate in the rest frame of the body allowed us to build a template for the optimal solution at each velocity. In the case of the bounded power, we use this template to study how the power can be modulated to vary the cruising velocity from start-up to supercritical velocities. A small acceleration $\dot{U}\ll\omega c$ is required to maintain the periodic assumption. In the bounded norm case, the intuition from the bounded power result is applied to find resonance points in the bounded norm case to minimise the forward wave and thus increase efficiency.

One point to note is that this work is not restricted to a fluid dynamics setting. The present work could also be used as a template to study the optimal conditions associated with variations of the wave equation and similar PDEs such as the heat, Schr\"{o}dinger and KdV equations. 
It would be interesting to replace the Galilean transformation with a relativistic Lorentz transformation to allow for time dilation. One major difference will be that the critical and supercritical velocities will be disallowed due to the constraint that $U/c<1$.

These results are of course limited by the assumptions made in the shallow water limit, but they do provide some general guidance. In order to achieve maximal efficiency, the goal based off of the intuition built here is to minimise the wave that is being produced ahead of the body. To refine this intuition, we wish to extend to the case where the surface of the shallow water is two dimensional. The expectation is to see a symmetry on either side of the centreline that is parallel to the direction of motion, much like what is observed when a honeybee propels itself along the water surface \citep{Roh2019bee}. Working in the same set of velocity regimes, a similar template for optimal solutions in the two dimensional case can be used generate an updated description of the quasiperiodic movement between different velocities under the restrictions of $\dot{U}\ll\omega c$. 

We can also study the problems outside of the shallow limit to build a model for the subsurface dynamics and dispersive waves. Some work with potential flow as used by \citet{Benham2024propulsion} could be applied with the addition of a moving reference frame to operate at different velocities. We could increase the complexity further with the full Navier-Stokes equations. This will introduce the potential for vortices to be generated by the movement of the body. Similar to \citet{GAO_FENG_2011} we could then estimate the role that the momentum transferred to the vortices plays in propulsion as well as the viscous and/or turbulent dissipation which are properties not present in the current framework. We could then study the difference in propulsion of the body from waves and vortices (WDP and VDP). Similar to \citet{Steinmann2021} we could estimate the values of the different forces at play to understand what dominates in our set-up at different scales. 

Unlike what is done by \citet{Benham2024propulsion}, the work presented in this paper only considers the raft as a pressure region. An interesting study would be to couple this work to a rigid or flexible beam and investigate the optimal solution for WDP in this case both theoretically and experimentally. The example of the step function sheds some light onto the type of optimal pressure distribution that may be required. Similarly, we could also look at optimal solutions for pressure distributions used in the literature like in the study of air-cushion vehicles \citep{Doctors_1972}. 

On our way to considering more physical setups, we can broaden our horizon past studying a periodic source. This will allow the implementation of multiple bodies on the fluid surface as well as the study of different incident waves on a single body. Removal of the assumption of a periodically oscillating body will permit the body to interact with external waves as well as allowing us to generate a more complete picture of accelerating between different velocities. We can then study an optimal control problem that could apply the wave-riding and wave-passing that ducklings demonstrate as described by \citet{Yuan2021ducks}. The interactions between the body and the waves will have to be studied in more detail since the body is always riding the waves produced by another. We could also use this version of the problem to investigate ways of optimising the distance between bodies to keep them stationary. 

This paper has helped us mathematically validate the key elements such as reducing the forward wave to maximise efficiency as well as build an understanding on the representation of WDP in the different velocity regimes. What is clear is there is plenty of more work to do, but this is a beginning that builds physical intuition in a simplified case that presents the opportunity for analytical as well as numerical study.

\section*{Funding}

{This work was partially funded by the European Research Council (ERC) under the EU Horizon 2020 research and innovation program (Grant Agreement 833125-HIGHWAVE)}

\vspace{1cm}

\noindent A github repository of example code used in this paper can be found at 
\url{https://github.com/daireodonovan/Optimal-locomotion-using-self-generated-waves.git}

\appendix

\section{Derivations of the shallow water wave equation, thrust and power}\label{qp_sec:derivations}

\subsection{Shallow water wave equation}\label{appendix_derive_shallow_wave_eq}

Beginning with the Euler equations, we will work in a one dimensional shallow water set up where the fluid is of height $H$ with characteristic length scale $L_{w}$. Through the use of the kinematic, dynamic and impermeable base boundary conditions we have a set of equations that can be solved. A pressure source perturbs the waves such that the dynamic boundary condition is,
\begin{equation}\label{ap_eq:dynamic_bc}
    p = P(x,t): \,\,\,  z = H + h(x,t),
\end{equation}
where $h(x,t)\ll H$ is the height of the resulting wave from the perturbation. We can begin by solving the rest case where $p=p_{a}$ (atmospheric pressure) followed by a perturbation with a pressure $p=\epsilon P(x,t)$, where $\epsilon = H/L_{w} \ll 1$ is the aspect ratio. The system of equations is taken to the leading order and the resulting horizontal momentum equation can be substituted into the vertically integrated conservation of mass equation to arrive at, 
\begin{align}\label{ap_eq:dimensional_wave}
    \frac{\partial^2h}{\partial x^2} - \frac{1}{c^2}\frac{\partial^2h}{\partial t^2} = -\frac{1}{\rho g}Q(x,t), && -\infty < x < \infty.
\end{align}

\subsection{Time-averaged thrust}\label{ap:thrust_derive}

We begin by manipulating the wave equation (in the subcritical moving frame) through multiplication by $\hat{h}_{\tilde{x}}$ and integrating, such that,
\begin{equation}\label{ap_eq:thrust_start}
	\int^{+}_{-}\left(h_{\tilde{t}\tilde{t}}-2vh_{\tilde{x}\tilde{t}}-(1-v^2)h_{\tilde{x}\tilde{x}}\right)h_{\tilde{x}}\,\mathrm{d}\tilde{x} = \int^{+}_{-} Qh_{\tilde{x}}\,\mathrm{d}\tilde{x},
\end{equation}
where we assume a periodic source such that the wave-field is also periodic,
\begin{equation}
    h = \mathrm{Re}\left(\hat{h}(\tilde{x})e^{i\tilde{t}}\right).    
\end{equation}
Substituting this in and taking the time average, we note that the second term on the left of \eqref{ap_eq:thrust_start} does not contribute to the thrust after time-averaging. This results in,
\begin{equation}\label{ap_eq:thrust_h_to_sub}
	\frac{1}{4}\left(v^2-1\right)\Big[\hat{h}^{'}\hat{h}^{'*}\Big]^{+}_{-}-\frac{1}{4}\Big[\hat{h}\hat{h}^{*}\Big]^{+}_{-} = \frac{1}{4}\int^{\frac{l}{2}}_{-\frac{l}{2}}\left(\hat{Q}^{*}\hat{h}^{'} + \hat{Q}\hat{h}^{'*}\right)\,\mathrm{d}\tilde{x}.
\end{equation}
Applying the boundary conditions on the right and left hand side of the body ($+$ and $-$ respectively)
\begin{align}\label{ap_eq:subcrit_bcs}
	\hat{h}^{'}(+) = -\frac{i}{1-v}\hat{h}(+),  &&\hat{h}^{'}(-) = \frac{i}{1+v}\hat{h}(-),
\end{align}
results in 
\begin{equation}
	\frac{1}{4}\left(\frac{v^2 - 1}{(1-v)^2}\big|\hat{h}\big|^2\Bigg|_{+} - \frac{v^2-1}{(1+v)^2}\big|\hat{h}\big|^2\Bigg|_{-} - \big|\hat{h}\big|^2\Bigg|_{+} + \big|\hat{h}\big|^2\Bigg|_{-}\right) = \langle\hat{Q},\hat{h}'\rangle ,
\end{equation}
\begin{equation}
	\implies -\frac{1}{2} \left(\frac{1}{1-v}\big|\hat{h}\big|^2\Bigg|_{+} - \frac{1}{1+v}\big|\hat{h}\big|^2\Bigg|_{-}\right) = -\frac{1}{2}\Big[k|\hat{h}|^2\Big]^{+}_{-} = \langle\hat{Q},\hat{h}'\rangle ,
\end{equation}
where we have used the notation defined in \eqref{eq:inner_product_defn}.
This can be defined as the thrust because the left hand side is equivalent to the radiation stress, with a difference in fore-aft amplitude. The $k$ acts as a correction to the wavenumber due to the waves getting Doppler shifted in the moving frame. Hence, we define the force as,
\begin{equation}
	\bar{F}_T \coloneq \langle\hat{Q},\hat{h}'\rangle .
\end{equation}

\subsection{Time-averaged power}\label{ap:power_derive}

Two routes can be taken to derive the time-averaged power. We could follow steps analogous to Appendix \ref{ap:thrust_derive} and instead of multiplying by $\hat{h}_{\tilde{x}}$ we instead multiply by $\hat{h}_{\tilde{t}}$. Here we will derive the power by taking the time derivative of the energy as it is commonly done.

Beginning with the dimensionless energy,
\begin{equation}
	E = \int^{(+)+vt}_{(-)+vt} \left(\frac{1}{2}h_{t}^2+\frac{1}{2}h_x^2 \right)\,\mathrm{d}x,
\end{equation}
we can transform the energy equation into the moving frame using the Galilean transformation in \eqref{eq:gal_transform},
\begin{equation}
	E = \int^{+}_{-} \left(\frac{1}{2}h_{\tilde{t}}^2+\frac{1}{2}(1+v^2)h_{\tilde{x}}^2 - vh_{\tilde{x}}h_{\tilde{t}} \right)\,\mathrm{d}\tilde{x}.
\end{equation}
Taking the time derivative will give us power,
\begin{equation}
	\frac{\mathrm{d}E}{\mathrm{d}\tilde{t}} = \int^{+}_{-}\Big((h_{\tilde{t}}-vh_{\tilde{x}})h_{\tilde{t}\tilde{t}} + (1+v^2)h_{\tilde{x}\tilde{t}}h_{\tilde{x}} - vh_{\tilde{t}}h_{\tilde{x}\tilde{t}}\Big)\,\mathrm{d}\tilde{x}.
\end{equation}
We can use the wave equation in the moving frame to substitute in for $h_{\tilde{t}\tilde{t}}$,
\begin{equation}
	\frac{\mathrm{d}E}{\mathrm{d}\tilde{t}}= \int^{+}_{-}\Big[Qh_{\tilde{t}}-vQh_{\tilde{x}} +(1-v^2)(h_{\tilde{x}}h_{\tilde{x}\tilde{t}} + h_{\tilde{t}}h_{\tilde{x}\tilde{x}}) - v(1-v^2)h_{\tilde{x}}h_{\tilde{x}\tilde{x}} + vh_{\tilde{t}}h_{\tilde{x}\tilde{t}} \Big] \,\mathrm{d}\tilde{x}.
\end{equation}
The first two elements of the integral will be left as they are because they include the control function $Q$, but the rest of the terms can be written in terms of derivatives with respect to $\tilde{x}$ and result in boundary terms
\begin{equation}
    \frac{\mathrm{d}E}{\mathrm{d}\tilde{t}}= \int^{+}_{-}\Big(Qh_{\tilde{t}}-vQh_{\tilde{x}}\Big)\,\mathrm{d}\tilde{x} +(1-v^2)\Big[h_{\tilde{x}}h_{\tilde{t}} \Big]^{+}_{-} - \frac{1}{2}v(1-v^2)\Big[h_{\tilde{x}}^2\Big]^{+}_{-} + \frac{1}{2}v\Big[h_{\tilde{t}}^2\Big]^{+}_{-}.
\end{equation}
We are working with periodic waves and if we take the time-average over a period, the result is $E(T)-E(0)=0$. Thus we make the equivalence relation between the power radiated by the waves and the power injected by the source,
\begin{equation}\label{ap_power_equality}
    -(1-v^2)\Big[\overline{h_{\tilde{x}}h_{\tilde{t}}} \Big]^{+}_{-} + \frac{1}{2}v(1-v^2)\Big[\overline{h_{\tilde{x}}^2}\Big]^{+}_{-} - \frac{1}{2}v\Big[\overline{h_{\tilde{t}}^2}\Big]^{+}_{-} = \int^{+}_{-}\Big(\overline{Qh_{\tilde{t}}-vQh_{\tilde{x}}}\Big)\,\mathrm{d}\tilde{x}.
\end{equation}
We can see that the second term on the right hand side is precisely the time-averaged thrust as derived in \ref{ap:thrust_derive}. Using the periodic assumption to decompose into spatial and temporal parts, we write the time-averaged power injected as, 
\begin{equation}
	\overline{\textrm{Pow}} = \langle\hat{Q},i\hat{h}\rangle - v \langle\hat{Q},\hat{h}'\rangle .
\end{equation} 
 Motivated by the thrust being written as the difference in fore-aft amplitude squared, we can expand out the left hand side of \eqref{ap_power_equality} in a similar way to what is done in \ref{ap:thrust_derive} to show,
 \begin{equation}\label{ap_eq:power_sum_h}
     \frac{1}{2}\left(\frac{1}{1-v}|\hat{h}_{+}|^2 + \frac{1}{1+v}|\hat{h}_{-}|^2\right) = \langle\hat{Q},i\hat{h}\rangle - v \langle\hat{Q},\hat{h}'\rangle .
 \end{equation}
Hence, the power is the sum of the fore-aft amplitude squared.

\section{Variational Calculus Approach}

\subsection{Case 1 - Bounded norm variational calculus}\label{norm_appendix}

The time-averaged thrust force from \eqref{eq:v0_thrust_with_quarter} is used with $\hat{h}(\tilde{x})$ \eqref{eq:sub_h_sol} substituted in for the subcritical case. This will simplify the following steps in the analytical optimisation approach where the thrust is given by,
\begin{equation}\label{ap_eq:simplified_thrust}
\begin{gathered}
    \bar{F}_T = \frac{1}{4}\Bigg(-\frac{1}{2(1-v)}\int^{\frac{l}{2}}_{-\frac{l}{2}}\hat{Q}(\tilde{x})\int^{\frac{l}{2}}_{-\frac{l}{2}}\hat{Q}^{*}(X)e^{\frac{i(\tilde{x}-X)}{1-v}}\,\mathrm{d}X\,\mathrm{d}\tilde{x} 
    \\
    + \frac{1}{2(1+v)}\int^{\frac{l}{2}}_{-\frac{l}{2}}\hat{Q}(\tilde{x})\int^{\frac{l}{2}}_{-\frac{l}{2}}\hat{Q}^{*}(X)e^{\frac{i(X-\tilde{x})}{1+v}}\,\mathrm{d}X\,\mathrm{d}\tilde{x}\Bigg).
\end{gathered}
\end{equation}
The action $f(\hat{Q})$ can be written with the thrust as the objective and the norm of the control function $\hat{Q}(\tilde{x})$ as the constraint
\begin{equation}\label{ap_eq:norm_action_expanded}
    \begin{gathered}
        f(\hat{Q}) = \frac{1}{4}\Bigg(-\frac{1}{2(1-v)}\int^{\frac{l}{2}}_{-\frac{l}{2}}\hat{Q}(\tilde{x})\int^{\frac{l}{2}}_{-\frac{l}{2}}\hat{Q}^{*}(X)e^{\frac{i(\tilde{x}-X)}{1-v}}\,\mathrm{d}X\,\mathrm{d}\tilde{x} 
        \\
        +\frac{1}{2(1+v)}\int^{\frac{l}{2}}_{-\frac{l}{2}}\hat{Q}(\tilde{x})\int^{\frac{l}{2}}_{-\frac{l}{2}}\hat{Q}^{*}(X)e^{\frac{i(X-\tilde{x})}{1+v}}\,\mathrm{d}X\,\mathrm{d}\tilde{x}\Bigg)
        +\frac{\lambda}{4}\left(\int^{\frac{l}{2}}_{-\frac{l}{2}}\hat{Q}^{*}\hat{Q}\,\mathrm{d}\tilde{x}\right),
    \end{gathered}
\end{equation}
where $\lambda$ is the Lagrange multiplier. 
Consider a perturbation $\epsilon \alpha(\tilde{x})$ to the function $\hat{Q}(\tilde{x})$. Variation of the action is defined as 
\begin{equation}\label{ap_eq:perturb_defn}
    \delta f = \lim_{\epsilon \rightarrow0}\frac{f\left(\hat{Q}(\tilde{x}) + \epsilon\alpha(\tilde{x})\right) - f\left(\hat{Q}(\tilde{x})\right)}{\epsilon}.
\end{equation}
The perturbed action is
\begin{equation}\label{ap_eq:norm_action_alpha}
    \begin{gathered}
        \delta f(\hat{Q}) = \frac{1}{4}\Bigg(-\frac{1}{2(1-v)}\int^{\frac{l}{2}}_{-\frac{l}{2}}\alpha(\tilde{x})\int^{\frac{l}{2}}_{-\frac{l}{2}}\hat{Q}^{*}(X)e^{\frac{i(\tilde{x}-X)}{1-v}}\,\mathrm{d}X\,\mathrm{d}\tilde{x} 
        \\
        + \frac{1}{2(1+v)}\int^{\frac{l}{2}}_{-\frac{l}{2}}\alpha(\tilde{x})\int^{\frac{l}{2}}_{-\frac{l}{2}}\hat{Q}^{*}(X)e^{\frac{i(X-\tilde{x})}{1+v}}\,\mathrm{d}X\,\mathrm{d}\tilde{x}\Bigg)
        +\frac{\lambda}{4}\left(\int^{\frac{l}{2}}_{-\frac{l}{2}}\alpha(\tilde{x})\,\hat{Q}^{*}(\tilde{x})\,\mathrm{d}\tilde{x}\right) + c.c.,
    \end{gathered}
\end{equation}
where $\textrm{c.c.}$ denotes the complex conjugate. 

When the variation is set to zero, the integrand of the $\tilde{x}$ integral can be interpreted as equalling to zero. The expression is further reduced by only taking the coefficients of $\alpha(\tilde{x})$ since the coefficients of $\alpha^{*}(\tilde{x})$ will also be zero since it is just the complex conjugate. Taking the complex conjugate of the results gives, 
\begin{equation}\label{ap_eq:norm_set_to_zero}
    -\frac{1}{2(1-v)}\int^{\frac{l}{2}}_{-\frac{l}{2}}\hat{Q}(X)e^{\frac{i(X-\tilde{x})}{1-v}}\,\mathrm{d}X + \frac{1}{2(1+v)}\int^{\frac{l}{2}}_{-\frac{l}{2}}\hat{Q}(X)e^{\frac{i(\tilde{x}-X)}{1+v}}\,\mathrm{d}X + \lambda \hat{Q}(\tilde{x}) = 0.
\end{equation}
The optimal condition can be found through operating on \eqref{ap_eq:norm_set_to_zero} with $\mathcal{L}_v = (1-v^2)\partial_{\tilde{x}\tilde{x}} + 2iv \partial_{\tilde{x}} + \mathbb{I}$. This has a null space proportional to $e^{\frac{i\tilde{x}}{1+v}}$ and $e^{-\frac{i\tilde{x}}{1-v}}$ as shown previously. Therefore, application of $\mathcal{L}_v$ on \eqref{ap_eq:norm_set_to_zero} results in
\begin{equation}\label{ap_eq:norm_cond}
    \mathcal{L}_v \hat{Q} = 0.
\end{equation}
This equation is easily solved and it tells us that $\hat{Q}(\tilde{x})$ is composed of the sum of two wave-like parts
\begin{equation}\label{ap_eq:norm_A_B_equation}
    \hat{Q}(\tilde{x}) = D_{\mathrm{L}}e^{\frac{i\tilde{x}}{1+v}} + D_{\mathrm{R}}e^{-\frac{i\tilde{x}}{1-v}},
\end{equation}
where the constants $D_{\mathrm{L}},D_{\mathrm{R}} \in \mathbb{C}$. This information can be substituted back into \eqref{ap_eq:norm_set_to_zero}, where the coefficients to $e^{\pm\frac{i\tilde{x}}{1\pm v}}$ are matched to generate two equations. Another equation is needed since there are three unknowns, $D_{\mathrm{L}}$, $D_{\mathrm{R}}$ and $\lambda$. The norm constraint in \eqref{eq:v0_norm_expression} is therefore bounded to an arbitrary value (we will choose 1) to get another equation to make the following system solvable,
\begin{subequations}\label{ap_eq:system_for_norm_cst}
    \begin{gather}
        \frac{1}{2(1+v)}\left(D_{\mathrm{L}}\,l + D_{\mathrm{R}}(1-v^2)\sin\left(\frac{l}{1-v^2}\right)\right) = -\lambda D_{\mathrm{L}},\label{eq:nrm_sys_a}
	    \\
	   -\frac{1}{2(1-v)}\left(D_{\mathrm{L}}(1-v^2)\sin\left(\frac{l}{1-v^2}\right) + D_{\mathrm{R}}\,l\right) = -\lambda D_{\mathrm{R}},\label{eq:nrm_sys_b}
       \\
       |D_{\mathrm{L}}|^2\,l + |D_{\mathrm{R}}|^2\,l + D_{\mathrm{L}}^{*}D_{\mathrm{R}}(1-v^2)\sin\left(\frac{l}{1-v^2}\right) + D_{\mathrm{L}}^{*}D_{\mathrm{R}}(1-v^2)\sin\left(\frac{l}{1-v^2}\right) = 4.\label{eq:nrm_sys_c}
    \end{gather}
\end{subequations}
Equation \eqref{eq:nrm_sys_b} demonstrates the relation between the constants $D_{\mathrm{L}}$ and $D_{\mathrm{R}}$,
\begin{equation}\label{ap_eq:A-Brelation}
	D_{\mathrm{R}} = D_{\mathrm{L}}\frac{(1-v^2)}{2(1-v)\lambda-l}\sin\left(\frac{l}{1-v^2}\right).
\end{equation}
Since the Lagrange multiplier $\lambda$ is real by definition, $D_{\mathrm{R}}$ is multiplied by a real number. This means that $D_{\mathrm{L}}$ and $D_{\mathrm{R}}$ are in phase. Equation \eqref{eq:nrm_sys_a} is used along with \eqref{ap_eq:A-Brelation} to show that 
\begin{equation}\label{ap_eq:general_lambda_nrm}
	\lambda_{\pm} = \frac{vl \pm \sqrt{ l^2 - (1-v^2)^3s^2}}{2(1-v^2)},
\end{equation}
where $s=\sin\left(\frac{l}{1-v^2}\right)$. As mentioned already, $\lambda\in\mathbb{R}$, which means that $l^2 - (1-v^2)^3s^2\geq 0$ must hold. It can be shown that this is satisfied for all real $l,v >0$. Information on $D_{\mathrm{L}}$ and therefore $D_{\mathrm{R}}$ can be found using \eqref{eq:nrm_sys_c},
\begin{equation}\label{ap_eq:A_nrm_result}
    |D_{\mathrm{L}}|^2 = 4\left(l+\frac{l(1-v^2)^2}{\left[2\lambda(1-v)-l\right]^2}\sin^2\left(\frac{l}{1-v^2}\right)+\frac{2(1-v^2)^2}{2\lambda(1-v)-l}\sin^2\left(\frac{l}{1-v^2}\right)\right)^{-1}.
\end{equation}
Finally \eqref{ap_eq:A-Brelation} is used to get information on $D_{\mathrm{R}}$,
\begin{equation}\label{ap_eq:B_nrm_result}
	|D_{\mathrm{R}}|^2 = \frac{|D_{\mathrm{L}}|^2(1-v^2)^2}{\big(2(1-v)\lambda-l\big)^2}\sin^2\left(\frac{l}{1-v^2}\right).
\end{equation}
All that is recovered is information on the modulus squared of $D_{\mathrm{L}},D_{\mathrm{R}}\in\mathbb{C}$. Thus, there are infinite possible solutions for $\hat{Q}$ defined up to an arbitrary phase.

In the interest of studying efficiency, we can write the force and power in terms of the constants $D_{\mathrm{L}}$ and $D_{\mathrm{R}}$. The force is given by 
\begin{gather}
    \bar{F}_T = \frac{1}{4} \left(-\frac{1}{2(1-v)}\left|\int^{\frac{l}{2}}_{-\frac{l}{2}}\hat{Q}(X)e^{\frac{iX}{1-v}}\,\textrm{d}X\right|^2 + \frac{1}{2(1+v)}\left|\int^{\frac{l}{2}}_{-\frac{l}{2}}\hat{Q}(X)e^{-\frac{iX}{1+v}}\,\textrm{d}X\right|^2\right),\label{ap_eq:force_to_sub_csts}
    \\
    \implies \bar{F}_T= \frac{1}{4}\left(-2(1-v)\lambda^2|D_{\textrm{R}}|^2 + 2(1+v)\lambda^2|D_{\textrm{L}}|^2\right).
\end{gather}
We have used the result in \eqref{ap_eq:norm_A_B_equation} and substituted in for $\lambda\hat{Q}$ in \eqref{ap_eq:norm_set_to_zero} to compare coefficients,
\begin{equation}\label{ap_eq:DL_DR_to_numerics}
    \begin{aligned}
        \frac{1}{2(1+v)}\int^{\frac{l}{2}}_{-\frac{l}{2}}\hat{Q}(X)e^{\frac{i(\tilde{x}-X)}{1+v}}\,\mathrm{d}X = -\lambda D_{\mathrm{L}},
        && &&
        \frac{1}{2(1-v)}\int^{\frac{l}{2}}_{-\frac{l}{2}}\hat{Q}(X)e^{\frac{i(X-\tilde{x})}{1-v}}\,\mathrm{d}X= \lambda D_{\mathrm{R}},
    \end{aligned}
\end{equation}
and the modulus squared of \eqref{ap_eq:DL_DR_to_numerics} is taken to write the thrust in terms of $D_{\mathrm{L}},D_{\mathrm{R}}. $ Similarly the power can be found to be,
\begin{equation}
    \overline{\textrm{Pow}} = \frac{1}{4}\left(2(1-v)\lambda^2|D_{\textrm{R}}|^2 + 2(1+v)\lambda^2|D_{\textrm{L}}|^2\right).
\end{equation}

\subsection{Case 2 - Bounded power variational calculus}\label{power_appendix}

A similar approach will be taken to the bounded norm case. Beginning with the expression for the thrust in \eqref{ap_eq:simplified_thrust}, the constraint in this case will be the power. Similar to what was done with the thrust, the power is simplified by substituting in for $\hat{h}(\tilde{x})$
\begin{equation}\label{ap_eq:expanded_power}
    \begin{gathered}
    \overline{\mathrm{Pow}} = \frac{1}{4}\left(\frac{1}{2}\int^{\frac{l}{2}}_{-\frac{l}{2}}\hat{Q}(\tilde{x})\int^{\frac{l}{2}}_{-\frac{l}{2}}\hat{Q}^{*}(X)e^{\frac{i(X-\tilde{x})}{1+v}}\,\mathrm{d}X\,\mathrm{d}\tilde{x} + \frac{1}{2}\int^{\frac{l}{2}}_{-\frac{l}{2}}\hat{Q}(\tilde{x})\int^{\frac{l}{2}}_{-\frac{l}{2}}\hat{Q}^{*}(X)e^{\frac{i(\tilde{x}-X)}{1-v}}\,\mathrm{d}X\,\mathrm{d}\tilde{x}\right)
    \\ 
    - v\bar{F}_{T}.
    \end{gathered}
\end{equation}
The resulting action is 
\begin{equation}\label{ap_eq:power_action}
    \begin{gathered}
        f(\hat{Q}) = (1-\lambda v)\bar{F}_T 
        \\
        +\frac{\lambda}{4}\left(\frac{1}{2}\int^{\frac{l}{2}}_{-\frac{l}{2}}\hat{Q}(\tilde{x})\int^{\frac{l}{2}}_{-\frac{l}{2}}\hat{Q}^{*}(X)e^{\frac{i(X-\tilde{x})}{1+v}}\,\mathrm{d}X\,\mathrm{d}\tilde{x} + \frac{1}{2}\int^{\frac{l}{2}}_{-\frac{l}{2}}\hat{Q}(\tilde{x})\int^{\frac{l}{2}}_{-\frac{l}{2}}\hat{Q}^{*}(X)e^{\frac{i(\tilde{x}-X)}{1-v}}\,\mathrm{d}X\,\mathrm{d}\tilde{x}\right).
    \end{gathered}
\end{equation}
Perturbing the action as it was defined in \eqref{ap_eq:perturb_defn} and setting the variation to be zero results in
\begin{equation}\label{ap_eq:pwr_perturb_let_zero}
    \begin{gathered}
         \left(-\frac{1}{2(1-v)}\int^{\frac{l}{2}}_{-\frac{l}{2}}\hat{Q}^{*}(X)e^{\frac{i(\tilde{x}-X)}{1-v}}\,\mathrm{d}X + \frac{1}{2(1+v)}\int^{\frac{l}{2}}_{-\frac{l}{2}}\hat{Q}^{*}(X)e^{\frac{i(X-\tilde{x})}{1+v}}\,\mathrm{d}X \right)
        \\
        +\frac{\lambda}{1-\lambda v}\left(\frac{1}{2}\int^{\frac{l}{2}}_{-\frac{l}{2}}\hat{Q}^{*}(X)e^{\frac{i(\tilde{x}-X)}{1-v}}\,\mathrm{d}X + \frac{1}{2}\int^{\frac{l}{2}}_{-\frac{l}{2}}\hat{Q}^{*}(X)e^{\frac{i(X-\tilde{x})}{1+v}}\,\mathrm{d}X \right)=0.
    \end{gathered}
\end{equation}
The complex conjugate of \eqref{ap_eq:pwr_perturb_let_zero} is then taken to work in terms of $\hat{Q}$,
\begin{equation}\label{ap_eq:eval_prob}
    \begin{gathered}
        \left(-\frac{1}{2(1-v)}\int^{\frac{l}{2}}_{-\frac{l}{2}}\hat{Q}(X)e^{\frac{i(X-\tilde{x})}{1-v}}\,\mathrm{d}X + \frac{1}{2(1+v)}\int^{\frac{l}{2}}_{-\frac{l}{2}}\hat{Q}(X)e^{\frac{i(\tilde{x}-X)}{1+v}}\,\mathrm{d}X\right) 
        \\
        +\frac{\lambda}{1-\lambda v}\left(\frac{1}{2}\int^{\frac{l}{2}}_{-\frac{l}{2}}\hat{Q}(X)e^{\frac{i(X-\tilde{x})}{1-v}}\,\mathrm{d}X + \frac{1}{2}\int^{\frac{l}{2}}_{-\frac{l}{2}}\hat{Q}(X)e^{\frac{i(\tilde{x}-X)}{1+v}}\,\mathrm{d}X \right)=0.
    \end{gathered}
\end{equation}
The first bracket of equation \eqref{ap_eq:eval_prob} is related to the second bracket by differentiation, leaving an eigenvalue problem to solve
\begin{equation}\label{ap_eq:eval_prob2}
    \left(\partial_{\tilde{x}} + \frac{i\lambda}{1-\lambda v}\right)\left(-\frac{i}{2}\int^{\frac{l}{2}}_{-\frac{l}{2}}\hat{Q}(X)e^{\frac{i(X-\tilde{x})}{1-v}}\,\mathrm{d}X - \frac{i}{2}\int^{\frac{l}{2}}_{-\frac{l}{2}}\hat{Q}(X)e^{\frac{i(\tilde{x}-X)}{1+v}}\,\mathrm{d}X\right) = 0.
\end{equation}
Equation \eqref{ap_eq:eval_prob2} has solutions of the form $Ce^{-\frac{i\lambda}{1-\lambda v}\tilde{x}}$, such that
\begin{equation}\label{ap_eq:power_gen_sol}
    -\frac{i}{2}\int^{\frac{l}{2}}_{-\frac{l}{2}}\hat{Q}(X)e^{\frac{i(X-\tilde{x})}{1-v}}\,\mathrm{d}X - \frac{i}{2}\int^{\frac{l}{2}}_{-\frac{l}{2}}\hat{Q}(X)e^{\frac{i(\tilde{x}-X)}{1+v}}\,\mathrm{d}X = Ce^{-\frac{i\lambda}{1-\lambda v}\tilde{x}}.
\end{equation}
To find $\lambda$, the operator $\mathcal{L}_v = (1-v^2)\partial_{\tilde{x}\tilde{x}} + 2iv\,\partial_{\tilde{x}} + \mathbb{I}$ is applied to \eqref{ap_eq:power_gen_sol}. All terms proportional to $e^{\frac{i\tilde{x}}{1+v}}$ and $e^{-\frac{i\tilde{x}}{1-v}}$ are part of the null space of $\mathcal{L}_v\hat{h}(\tilde{x})$ and hence vanish. As a result we have,
\begin{equation}\label{ap_eq:to_find_lambda}
    \mathcal{L}_ve^{-\frac{i\lambda}{1-\lambda v}\tilde{x}} = 0,
\end{equation}
and two solutions for $\lambda$ are recovered,
\begin{align}
   \lambda_1 = -1 &&  \lambda_2 = 1.
\end{align}
There are two possible solutions to the eigenvalue problem,
\begin{align}
        C_{\mathrm{L}}\,e^{\frac{i\tilde{x}}{1+v}} + C_{\mathrm{R}}\,e^{-\frac{i\tilde{x}}{1-v}} = C\,e^{\frac{i\tilde{x}}{1+v}} & & & \textrm{for  } \lambda_1,
        \\ 
        C_{\mathrm{L}}\,e^{\frac{i\tilde{x}}{1+v}} + C_{\mathrm{R}}\,e^{-\frac{i\tilde{x}}{1-v}} = C\,e^{-\frac{i\tilde{x}}{1-v}} & & & \textrm{for}\,\, \lambda_2 .
\end{align}
Each solution has one constant set to zero and the other non-zero. This corresponds to a purely left or right travelling wave. Expressions for $C_{\mathrm{L}}$ and $C_{\mathrm{R}}$ are found by matching wave coefficients in \eqref{ap_eq:power_gen_sol},
\begin{align}\label{ap_eq:equated_coeffs}
    C_{\mathrm{L}} = -\frac{i}{2}\int^{\frac{l}{2}}_{-\frac{l}{2}}\hat{Q}(X)e^{-\frac{iX}{1+v}}\,\mathrm{d}X, & & C_{\mathrm{R}} = -\frac{i}{2}\int^{\frac{l}{2}}_{-\frac{l}{2}}\hat{Q}(X)e^{\frac{iX}{1-v}}\,\mathrm{d}X.
\end{align}
To find conditions for $C_{\mathrm{L}}$ and $C_{\mathrm{R}}$ to satisfy, the power \eqref{ap_eq:power_sum_h} is used,
\begin{equation}\label{ap_eq:constants_equation}
    \frac{2}{(1+v)}\left|C_{\mathrm{L}}\right|^2 + \frac{2}{(1-v)}\left|C_{\mathrm{R}}\right|^2 = 4,
\end{equation}
where \eqref{ap_eq:equated_coeffs} is used to substitute in $C_{\mathrm{L}}$ and $C_{\mathrm{R}}$.
To maximise thrust in the positive direction, we will choose a purely left travelling wave ($\lambda_1$). This choice sets $C_{\mathrm{L}} = C$ and $C_{\mathrm{R}} = 0$,
\begin{align}\label{ap_eq:amplitude_cond}
    \left|-\frac{i}{2}\int^{\frac{l}{2}}_{-\frac{l}{2}}\hat{Q}(X)e^{-\frac{iX}{1+v}}\,\mathrm{d}X\right|^2 = 2(1+v), & & \left|-\frac{i}{2}\int^{\frac{l}{2}}_{-\frac{l}{2}}\hat{Q}(X)e^{\frac{iX}{1-v}}\,\mathrm{d}X\right|^2=0.
\end{align}
This is the condition that has to be met for the optimal solution comprising of a left travelling wave only. Similarly, the thrust can be written in terms of $C_{\mathrm{L}}$ and $C_{\mathrm{R}}$ and when the conditions in \eqref{ap_eq:equated_coeffs} are substituted into \eqref{ap_eq:force_to_sub_csts}, the optimum thrust for a velocity $v$ is recovered
\begin{equation}\label{ap_eq:optimum_thrust}
    \bar{F}_T = \frac{1}{4}\left(\frac{2}{1+v}\left|C_{\mathrm{L}}\right|^2 - \frac{2}{1-v}\left|C_{\mathrm{R}}\right|^2\right) = 1.
\end{equation}

\bibliographystyle{jfm}
\bibliography{bibfile}

\end{document}